\documentclass[iop,usenatbib]{emulateapj}

\usepackage{journals}
\usepackage{color}
\usepackage{graphicx}
\usepackage{multirow}
\usepackage{amsmath}
\definecolor{urlcolour}{rgb}{0,0,0}
\definecolor{citecolour}{rgb}{0,0,0}
\definecolor{linkcolour}{rgb}{0,0,0}
\usepackage{hyperref}
\hypersetup{colorlinks,breaklinks, urlcolor=urlcolour, linkcolor=linkcolour, citecolor=citecolour}

\shorttitle{RCSED -- Reference Catalog of Galaxy SEDs}
\shortauthors{Chilingarian et al.}

\begin{document}
\title{RCSED -- A Value-Added Reference Catalog of 
Spectral Energy Distributions of 800,299 Galaxies in 11 Ultraviolet, Optical, and
Near-Infrared Bands: Morphologies, Colors, Ionized Gas and Stellar
Populations Properties}
\author{Igor V. Chilingarian\altaffilmark{1,2,a},
Ivan Yu. Zolotukhin\altaffilmark{3,2,4,b},
Ivan Yu. Katkov\altaffilmark{2}, 
Anne-Laure Melchior\altaffilmark{5}, 
Evgeniy V. Rubtsov\altaffilmark{2,6},
Kirill A. Grishin\altaffilmark{6}}

\affil{$^1$ Smithsonian Astrophysical Observatory, 60 Garden St. MS09, Cambridge, MA, 02138, USA}
\affil{$^2$ Sternberg Astronomical Institute, M.V. Lomonosov Moscow State University, 13 Universitetsky prospect, Moscow, 119991, Russia}
\affil{$^3$ Universit\'e de Toulouse; UPS-OMP, IRAP, 9 avenue du Colonel Roche, BP 44346, F-31028 Toulouse Cedex 4, France}
\affil{$^4$ Special Astrophysical Observatory of the Russian AS, Nizhnij Arkhyz 369167, Russia}
\affil{$^5$ Sorbonne Universit\'es, UPMC Univ. Paris 6, Observatoire de Paris, PSL Research University, CNRS,UMR 8112, LERMA, Paris, France}
\affil{$^6$ Department of Physics, M.V. Lomonosov Moscow State University, 1, Leninskie Gory, Moscow, Russia, 119991}
\email{$^{A}$E-mail: igor.chilingarian@cfa.harvard.edu}
\email{$^{B}$E-mail: ivan.zolotukhin@irap.omp.eu}

\begin{abstract} 
We present RCSED\footnote{The
data tables and other supporting technical information are available 
at the project web-site: \url{http://rcsed.sai.msu.ru/}},
the value-added Reference Catalog of Spectral Energy
Distributions of galaxies, which contains homogenized spectrophotometric
data for 800,299 low and intermediate redshift galaxies ($0.007<z<0.6$)
selected from the Sloan Digital Sky Survey spectroscopic sample.  Accessible
from the Virtual Observatory (VO) and complemented with
detailed information on galaxy properties obtained with the state-of-the-art
data analysis, RCSED enables direct studies of galaxy formation and
evolution during the last 5~Gyr.  We provide tabulated color transformations
for galaxies of different morphologies and luminosities and analytic
expressions for the red sequence shape in different colors.  RCSED comprises
integrated $k$-corrected photometry in up-to 11 ultraviolet, optical, and
near-infrared bands published by the GALEX, SDSS, and UKIDSS wide-field imaging
surveys; results of the stellar population fitting of SDSS spectra including
best-fitting templates, velocity dispersions, parameterized star formation
histories, and stellar metallicities computed for instantaneous starburst
and exponentially declining star formation models; parametric and
non-parametric emission line fluxes and profiles; and gas phase
metallicities.  We link RCSED to the Galaxy Zoo morphological classification
and galaxy bulge+disk decomposition results by Simard et al.  We construct
the color--magnitude, Faber--Jackson, mass--metallicity relations, compare
them with the literature and discuss systematic errors of galaxy properties
presented in our catalog.  RCSED is accessible from the project web-site and
via VO simple spectrum access and table access services using VO compliant
applications.  We describe several SQL query examples against the
database.  Finally, we briefly discuss existing and future scientific
applications of RCSED and prospectives for the catalog extension to higher
redshifts and different wavelengths.  
\end{abstract}


\section{Introduction and Motivation}

During the last decade we witnessed a breakthrough in wide field imaging
surveys across the electromagnetic spectrum.  The new era started with the
Sloan Digital Sky Survey (SDSS) that used a 2.5-m telescope and covered over
11,600~sq.~deg.  of the sky in 5 optical photometric bands ($ugriz$) down to
the 22nd AB magnitude in its latest 7th legacy data release \citep{SDSS_DR7}.  It had
a spectroscopic follow-up survey that targeted over 1 million galaxies and
quasars and half a million stars down to the magnitude limit of
$r=17.77$~AB~mag.  Even though by the end of 2015, the data from SDSS and its
successors, SDSS-II, and SDSS-III were used in about 20,000 research
papers\footnote{According to NASA ADS, \url{http://ads.harvard.edu/}}, the
SDSS potential for scientific exploration remains far from exhaustion.

In the late 2000s, deep wide field surveys went beyond the optical spectral
domain.  The Galaxy Evolution Explorer (GALEX) satellite \citep{Martin+05}
provided nearly all-sky photometric coverage in two ultraviolet bands
centered at 154 and 228~nm down to the limiting magnitudes $AB = 20.5$~mag. 
The SDSS footprint area was observed by GALEX with 15 times longer exposure that
yielded a much deeper limit of $AB = 23.5$~mag.  The relatively small
telescope provided the spatial resolution of a couple of arcsec comparable
to the typical image quality level at ground-based facilities.

At the same time, a major effort was undertaken by the international team at
the 4-m United Kingdom Infrared Telescope UKIRT to survey a substantial area
of the sky largely overlapping with the SDSS footprint in 4 near-infrared
(NIR) bands ($YJHK$).  The Large Area Survey of the UKIRT Deep Sky Survey
(UKIDSS LAS, \citealp{Lawrence+07}) provides a sub-arcsecond resolution and
the flux limit comparable to that of SDSS in the optical domain.  It reaches
$AB \sim 21.2$~mag, 3--4~mag deeper than the first all-sky NIR survey 2MASS
\citep{Skrutskie+06}.

Numerous projects studied the entire SDSS spectroscopic sample of galaxies
by analyzing both absorption \citep[see e.g.][]{Kauffmann+03,GCBW06} and
emission lines \citep{Brinchmann+04,Tremonti+04,OSSY11,Oh+15} in SDSS
spectra (MPA-JHU and OSSY catalogs).  However, they did
not make use of any additional information beyond that available in the SDSS
database.

The first successful attempt of providing an added value to SDSS data was
done a decade ago in the ``New York University Value-Added Galaxy Catalog''
(NYU-VAGC) project \citep{Blanton+05}.  It was aimed at statistical studies
of galaxy properties and the large scale structure of the Universe and
included a compilation of information derived from photometry and
spectroscopy in one of the earlier SDSS data releases (that represents about
20\%\ of its final imaging footprint).  It also comprised positional
cross matches with 2MASS, far-infrared \emph{IRAS} point source catalog
\citep{Saunders+00}, the Faint Images of the Radio sky 20~cm survey FIRST
\citep{BWH95}, and additional data on galaxies from the 3rd Reference
Catalogue of Bright Galaxies \citep{deVaucouleurs+91} and the Two-Degree
Field Galaxy Redshift Survey \citep{Colless+01}.  Now, a decade after the
NYU-VAGC has been published, there is a sharp need to assemble a next
generation of a value-added galaxy catalog based on modern survey data that
were not available back then.

Here we present a new generation and a different flavor of a value-added
catalog of galaxies based on a combination of data from SDSS, GALEX, and
UKIDSS surveys that also includes comprehensive analysis of absorption and
emission lines in galaxy spectra.  Our main motivation is to use the synergy
provided by the joint panchromatic dataset for extragalactic astrophysics:
the optical domain is traditionally the best studied and there exist well
calibrated stellar population models; the UV fluxes are
sensitive to even small fractions of recently formed stars and therefore
contain valuable information on star formation histories; the near-IR band
is substantially less sensitive to the internal dust reddening and stellar
population ages, and therefore can provide good stellar mass estimates.  Our
mission is to build a reference multi-wavelength spectrophotometric dataset
and complement it with additional detailed information on galaxy properties
so that it will allow astronomers to study galaxy formation and evolution at
redshifts $z=0.0\textup{--}0.6$ in a transparent way with as little extra
manipulations as possible.

We aim to provide:
(i) the first homogeneous set of low redshift galaxy FUV-to-NIR spectral
energy distributions (SEDs) corrected to rest frame for hundreds of
thousands of objects;
(ii) the first photometric dataset containing rest frame aperture SEDs with
corresponding spectra and their stellar population analysis: velocity
dispersions, parameterized star formation histories;
(iii) consistent analysis of absorption and emission lines in
SDSS galaxy spectra including parametric and non-parametric emission 
line fitting performed using state-of-the-art stellar population models,
which cover a wider range of ages and metallicities and, therefore, help to
minimize the template mismatch;
(iv) easy and fully Virtual Observatory compliant data access mechanisms 
for our dataset and several third-party catalogs that include morphological 
and structural information for galaxies in our sample.

We started this project in 2009 by developing a new approach to convert
galaxy SEDs to the rest frame by calculating analytic approximations of
$k$-corrections in optical and NIR bands \citep{CMZ10}.  Then we extended
our algorithm to GALEX \emph{FUV} and \emph{NUV} bands and discovered a
universal 3-dimensional relation of \emph{NUV} and optical galaxy colors and
luminosities \citep{CZ12}.  Then, we fitted SDSS spectra using
state-of-the-art stellar population models, derived velocity dispersions and
stellar ages and metallicities, and provided our measurements to the project
that calibrated the fundamental plane of galaxies \citep{DD87} in SDSS by
vigorous statistical analysis \citep{SMZC13}. Our dataset also helped to
find and characterize massive compact early-type galaxies at intermediate
redshifts \citep{DCHG13,DHGC14}. Finally, we used a complex
set of selection criteria and discovered a large sample of previously
considered extremely rare compact elliptical galaxies \citep{CZ15}.

The paper is organized as follows: in \emph{Section~2} we describe the
construction of the catalog that includes cross-matching of the three
surveys, adding third-party catalogs, absorption and emission line
analysis of SDSS spectra; in \emph{Section~3} we discuss the photometric
properties of the sample and derive mean colors of galaxies of different
morphological types across the spectrum; in \emph{Section~4} we explore the
information derived from our spectral analysis; \emph{Section~5} contains
the description of the catalog access interfaces; \emph{Section~6}
provides the summary of our project; and \emph{Appendices} include some
technical details on the catalog construction, detailed description of
tables included in the database, and discussion of systematic uncertainties
of emission line measurements.

\section{Construction of the catalog}

\subsection{The input sample and data sources used.}
\label{sec_sample}

We compiled the photometric catalog by re-processing several publicly available datasets. 
Our core object list is the SDSS Data Release 7 \citep{SDSS_DR7} spectral sample of non-active galaxies (marked as ``GAL\_EM'' or ``GALAXY'' specclass in the SDSS database) in the redshift range $0.007 \le z < 0.6$. 
We provide the exact query that we used to select this sample in the SDSS CasJobs Data System\footnote{\url{http://skyserver.sdss3.org/CasJobs/}} in Appendix~\ref{sec_sql}.
The query executed in the DR7 CasJobs context returned 800,299 records. 
We deliberately excluded quasars and Seyfert-1 galaxies (specclass=``QSO'') because neither the $k$-correction technique, nor stellar population analysis algorithm supported that object type.
We used the output table as an input list for positional cross-matches against GALEX Data Release 6 \citep{Martin+05} and UKIDSS Data Release 10 \citep{Lawrence+07}.  

For the UKIDSS cross-match we queried the UKIDSS Large Area Survey catalog using the best match criterion within a 3~arcsec radius. 
In order to perform this query, we employed the WFCAM Science Archive\footnote{\url{http://surveys.roe.ac.uk/wsa/}} for the programmatic access to the International Virtual Observatory Alliance (IVOA) ConeSearch service with a multiple cone search (``multi-cone'') capability. 
The query returned 280,870 UKIDSS objects matching the galaxies from our input sample.
We used the {\sc stilts} software package \citep{Taylor06} in order to access the UKIDSS data and merge the tables.

Then we uploaded the input SDSS galaxy list to the GALEX CasJobs web interface\footnote{\url{http://galex.stsci.edu/casjobs/}} and searched best matches within 3 arcsec similarly to the UKIDSS cross-match. 
The query returned 485,996 GALEX objects.

\begin{figure}
\includegraphics[width=\hsize]{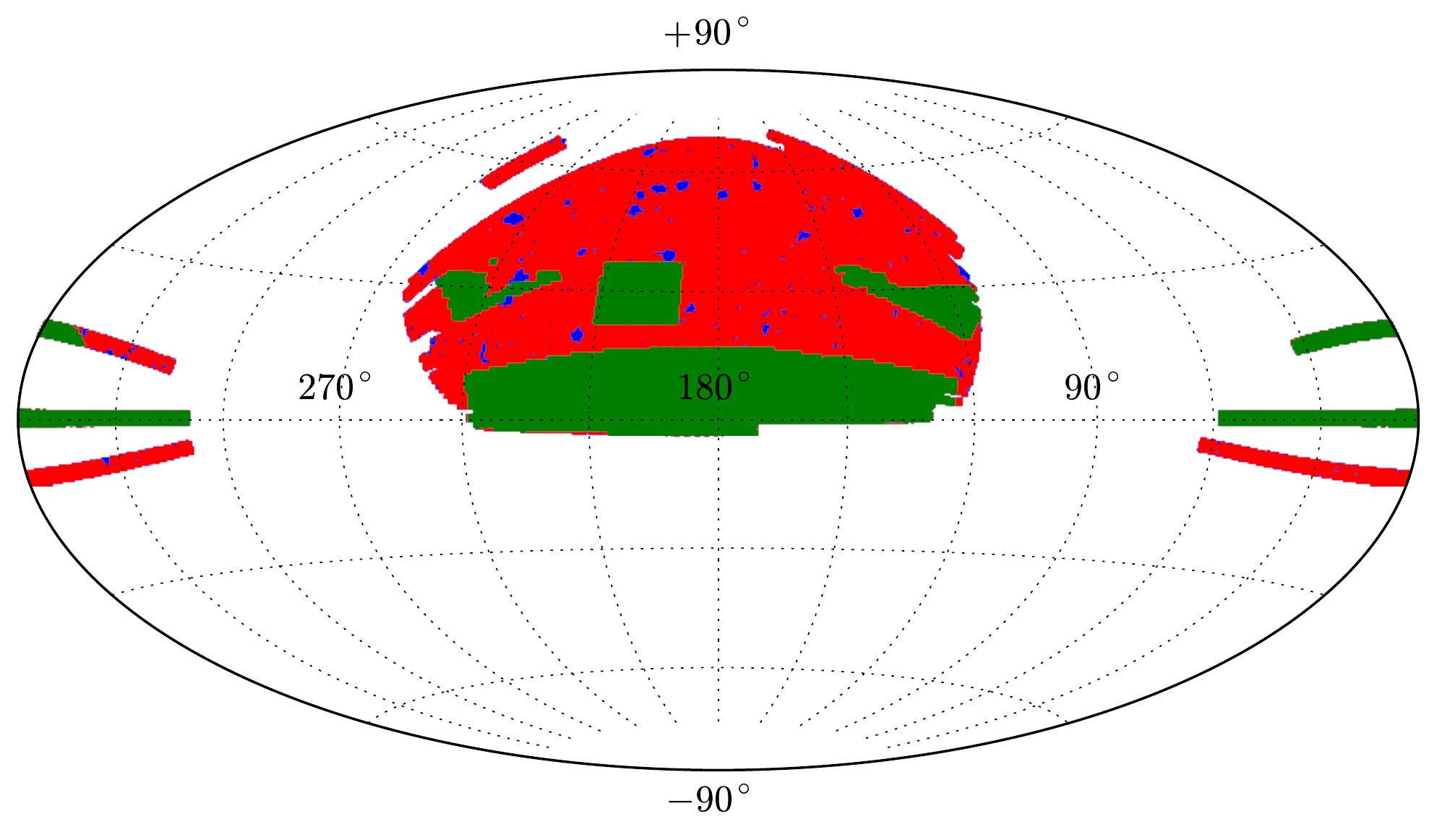}
\caption{A full sky aitoff projection in equatorial coordinates demonstrating the footprint of our
catalog. Green areas denote the availability of all three input photometric
datasets, SDSS, UKIDSS, and GALEX; red areas are for SDSS and GALEX; and blue
areas are for SDSS only. Note that we include all objects from the input datasets that have at least one flux measurement in them.
\label{fig_samplemap}}
\end{figure}

As a result of this selection procedure we compiled an input catalog of
800,299 spectroscopically confirmed SDSS galaxies, out of which 90,717
have 11 band photometry (two \textit{GALEX} \textit{FUV} and \textit{NUV}, 5
\textit{SDSS} $ugriz$ bands, 4 UKIDSS $YJHK$ bands), 163,709 have all UKIDSS
bands and at least one UV band, 582,534 have at least one additional
photometric band to SDSS bands. In Fig.~\ref{fig_samplemap} we present the
footprint of our catalog on the all-sky aitoff projection marking the
regions covered by all three wide field imaging surveys using different
colors. The statistics of galaxies measured in different photometric bands
is given in Table~\ref{tab_photnumb}.

\begin{deluxetable}{lr}
\tablecolumns{2}
\tablecaption{Number of objects in the combined sample with photometric
measurements available from three input photometric catalogs.
\label{tab_photnumb}}
\tablehead{
\colhead{Photometric bands} & \colhead{Number of galaxies}
}
\startdata
SDSS $ugriz$ &      799783\\
GALEX $FUV$ + $ugriz$ &      286570\\
GALEX $NUV$ + $ugriz$ &      469419\\
$FUV$ + $NUV$ + $ugriz$ &      270152\\
$ugriz$ + UKIDSS $Y$ &      270603\\
$ugriz$ + UKIDSS $J$ &      265316\\
$ugriz$ + UKIDSS $H$ &      272028\\
$ugriz$ + UKIDSS $K$ &      273050\\
$ugriz + YJHK$ &      250608\\
$NUV + ugriz + YJHK$ &      157531\\
all 11 bands &       90717
\enddata
\end{deluxetable}

Then we linked the following published datasets to our catalog in order to
contribute the spectrophotometric information with some of the most widely
used galaxy properties: (i) the results of the two-dimensional light
profile decomposition of SDSS galaxies by \citet{Simard+11} that include
structural properties of all objects in our catalog; (ii) the morphological
classification table from the citizen science ``Galaxy Zoo'' project
\citep{Lintott+08,Lintott+11} that provides a human eye classification of
well spatially resolved SDSS galaxies made by citizen scientists. 661,319
objects in our sample have 10 or more morphological classifications in the
Galaxy Zoo catalog ($nvote \ge 10$).

\subsection{The photometric catalog}

\subsubsection{Petrosian and aperture magnitudes}

All three photometric surveys used in our study provide extended source
photometry along with aperture measurements made in several different
aperture sizes (GALEX and UKIDSS).  

For the SED photometric analysis and construction of scaling relations
involving galaxy luminosity, we need total magnitudes.  For this purpose we
adopt \citet{Petrosian76} magnitudes available in SDSS and UKIDSS as
measurements which do not significantly depend on galaxy light profile
shapes conversely to SDSS \emph{modelmags} \citep[see discussion
in][]{CZ12}.  The GALEX catalog provides ``total'' magnitudes that are close
to Petrosian magnitudes for exponential surface brightness profiles (i.e. 
disc galaxies) and up-to 0.2~mag brighter for elliptical galaxies
\citep{Yasuda+01}.  However, given the average photometric uncertainty in
the GALEX $NUV$ fluxes of red galaxies of $0.3$~mag, we can neglect this
difference.

On the other hand, our parent sample of galaxies was derived from the SDSS
spectroscopic sample, and all SDSS DR7 spectra were obtained in circular
3-arcsec wide apertures.  Therefore, we need 3-arcsec aperture magnitudes in
order to make quantitative comparison of spectroscopic and photometric data. 
Hence, we computed aperture magnitudes for all GALEX and UKIDSS sources with available
aperture measurements by interpolating the flux to a 3-arcsec aperture, and used SDSS \emph{fibermags} for the optical SED part. 
To be noticed, that the spatial resolution of the GALEX survey in the $NUV$
band is about 5~arcsec, therefore 3-arcsec aperture magnitudes will be
slightly underestimated for small objects. For compact (point-like)
sources a 3-arcsec aperture $NUV$ magnitude can be underestimated by as 
much as 0.3~mag, however, such objects are very rare in the SDSS
DR7 galaxy sample. \citet{DCHG13,ZDGC15,Zahid+16} found a couple of
thousands compact sources in SDSS and SDSS-{\sc iii} BOSS, only a few 
hundreds of which were in SDSS DR7. We estimated a number of compact
galaxies in our sample by selecting the sources where the average difference
of aperture and Petrosian magnitudes in $ugriz$ bands was $<0.3$~mag: this
query returned 831 objects or $<0.1$\% of the sample.

We corrected the obtained sets of Petrosian and 3-arcsec aperture magnitudes
for the Galactic foreground extinction by using the $E(B-V)$ values computed
from the \citet{SFD98} extinction maps. Then, we computed $k$-corrections
for both sets of photometric points using the analytic approximations
presented in \citet{CMZ10} and updated for $GALEX$ bands in \citet{CZ12}.

\begin{figure}
\includegraphics[width=\hsize]{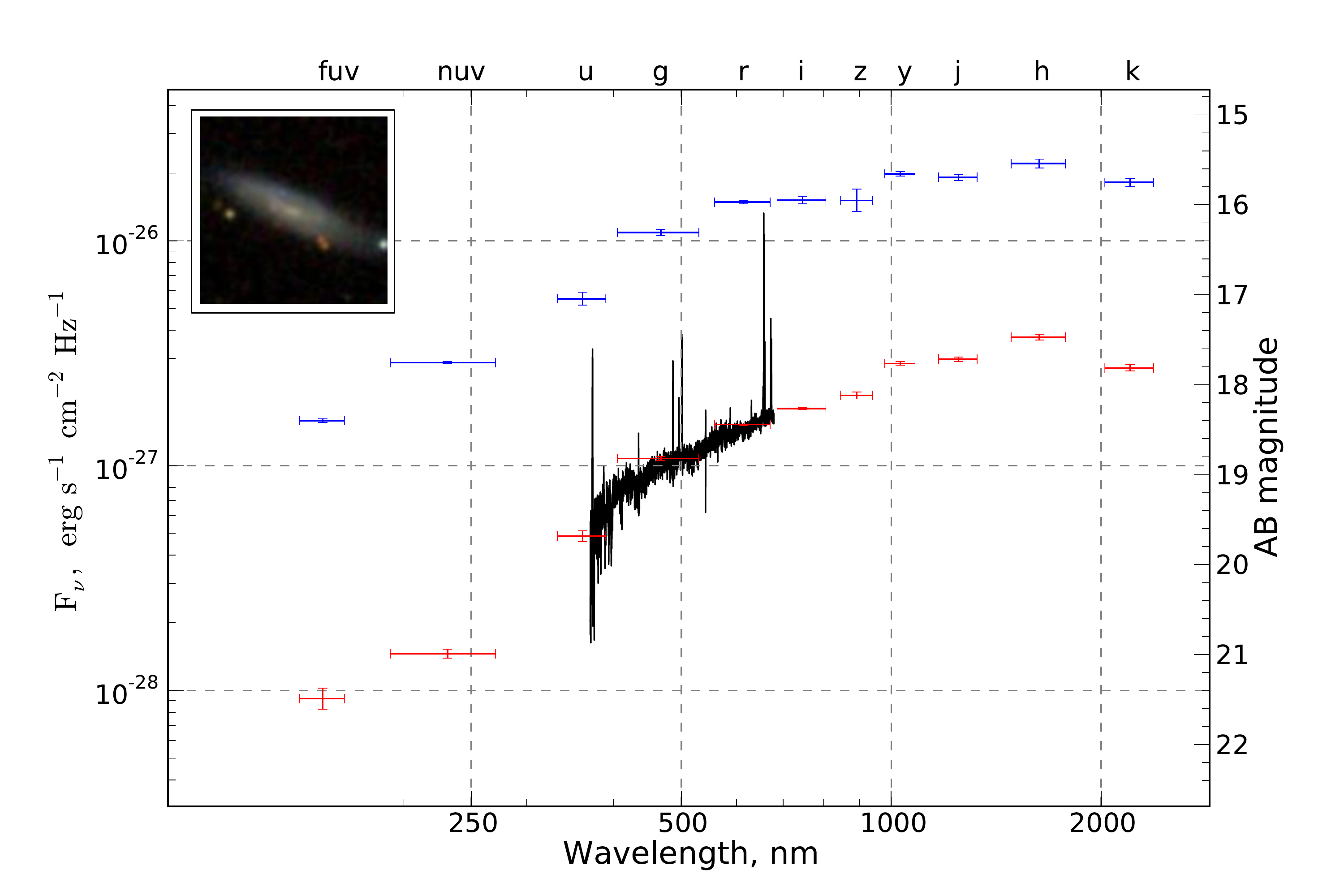}
\caption{Example of fully corrected SED in 11 bands for a late type spiral
galaxy at redshift 0.035. Blue and red symbols represent total (Petrosian) and 3-arcsec fiber
magnitudes correspondingly. The rest framed SDSS spectrum is overplotted and
demonstrates a typical excellent agreement with the corrected fiber magnitudes for
that galaxy. The inset shows an 36$\times$36~arcsec optical SDSS false color 
image.
\label{fig_sed}}
\end{figure}

In Fig.~\ref{fig_sed} we provide an example of a fully corrected SED for a
late type spiral galaxy ($z=0.035$) that has flux measurements in all 11
bands.  We show both total and fiber magnitudes and overplot an SDSS
spectrum with the wavelength axis converted into the rest frame and fluxes
converted into $AB$ magnitudes.  One can see a remarkable agreement between
the corrected photometric points and the observed spectral flux density, typical for our catalog.

\subsubsection{Correcting the SDSS--UKIDSS photometric offset}

An important problem of the UKIDSS photometric catalog of extended sources
is the observed spread of colors including optical SDSS and NIR UKIDSS
photometric measurements (e.g.  $g-J$ for red sequence galaxies).  We
detected this inconsistency in \citet{CMZ10} and applied an empirical
correction to UKIDSS magnitudes based on the assumption of continuous SEDs
of galaxies.  We computed $z-Y$ colors by interpolating over all other
available colors approximating the SED with a low order polynomial function. 
This approach, however, required the availability of the $Y$ band photometry
in the UKIDSS catalog.  We have analyzed the SDSS--UKIDSS Petrosian
magnitude offset amplitude for different galaxies and concluded, that it
originates from the surface brightness limitation imposed by relatively
short integration time in UKIDSS and by high and variable sky background
level in the NIR.  Therefore, Petrosian radii and magnitudes become
underestimated, and comparison of original UKIDSS extended source magnitudes
with SDSS and GALEX integrated photometry becomes impossible, because any
color including data from UKIDSS and another data source depends on the
galaxy surface brightness and size.

Here we propose a general and simple empirical solution.  We
exploit the UKIDSS Galactic Cluster Survey photometric catalog that includes
the $Z$ band photometry, convert it into SDSS $z$ with the available color
transformation \citep{HWLH06} for both Petrosian and 3-arcsec aperture
magnitudes, and compare it to actual SDSS $z$ band measurements from the
SDSS DR7 catalog for exactly the same objects.  It turns out, that (i)
the Petrosian magnitude difference $z_{\mathrm{SDSS,Petro}} -
z_{\mathrm{UKIDSS,Petro}}$ correlates with the galaxy mean surface
brightness; (ii) the fiber magnitude difference $z_{\mathrm{SDSS,fib}} -
z_{\mathrm{UKIDSS,3''}}$ is close to zero within 0.02~mag; (iii) differences
between Petrosian and fiber magnitudes in all UKIDSS photometric bands
($ZYJHK$) are almost identical that indicates virtually flat NIR color
profiles in most galaxies.  This suggests that the correction for UKIDSS
Petrosian magnitudes should be calculated as: $\Delta
(mag_{\mathrm{UKIDSS,Petro}}) = (z_{\mathrm{SDSS,fib}} -
z_{\mathrm{SDSS,Petro}}) - (Y_{\mathrm{UKIDSS,3''}} -
Y_{\mathrm{UKIDSS,Petro}})$.  This transformation adjusts the UKIDSS
integrated photometry in a way that the differences between the 3~arcsec and
Petrosian magnitudes of a galaxy in $z$ and $Y$ bands become equal. For
objects, where $Y$ magnitudes are not available in the UKIDSS survey, we use
the next available photometric band ($J$, $H$, or $K$).

In this fashion, we obtained fully corrected \emph{FUV-to-NIR} spectral
energy distributions converted into rest-frame magnitudes for a large sample
of galaxies in 3-arcsec apertures and integrated over entire galaxies.

\subsection{The spectral catalog: absorption lines}

\begin{figure}
\includegraphics[clip,trim={1cm 0 2.5cm 0},width=\hsize]{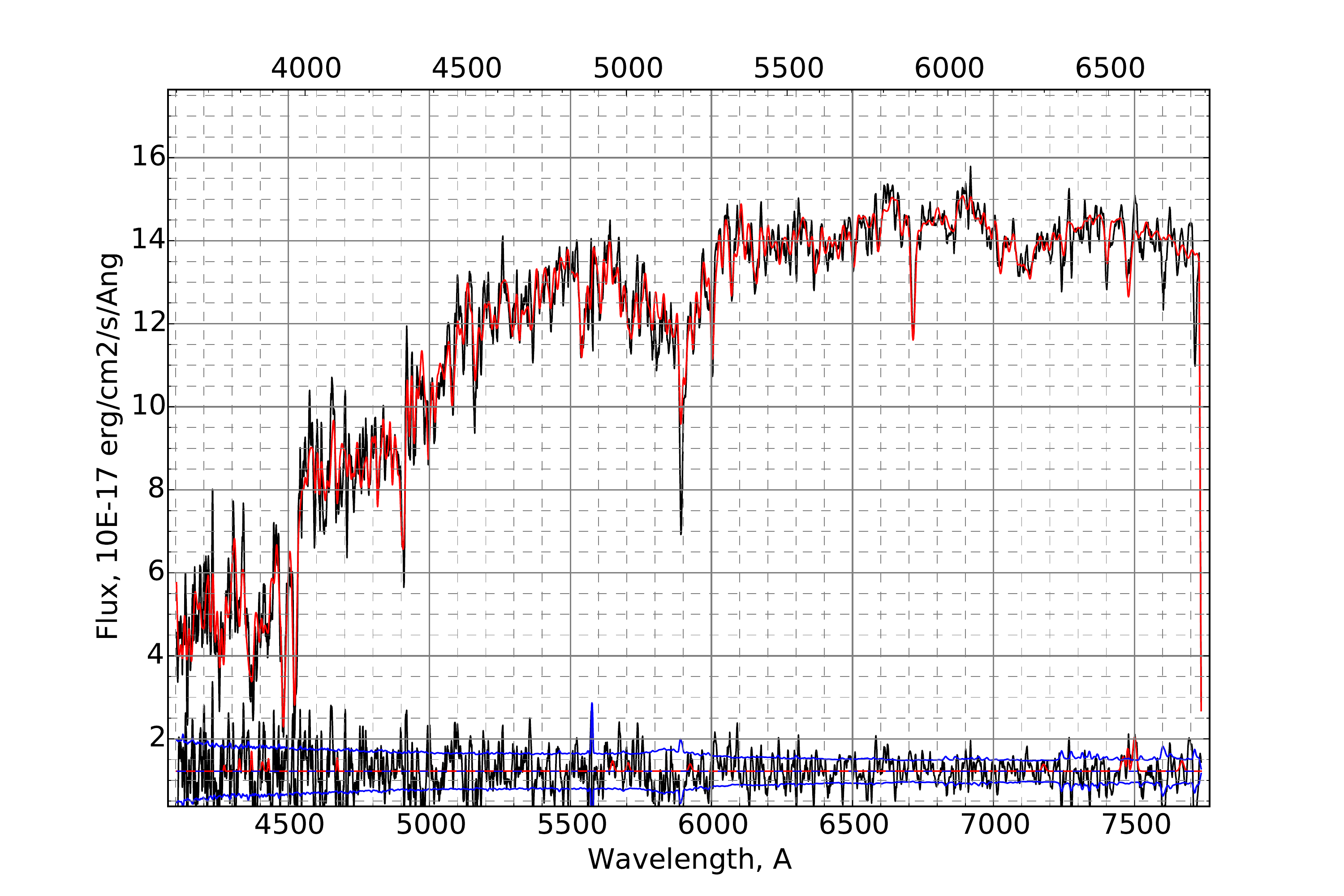}
\caption{Example of the {\sc nbursts} full spectrum fitting for an SDSS
spectrum of an early type galaxy. An observed galaxy spectrum is shown in
black, the best fitting template is in red, residuals are in blue. Regions
of emission lines excluded from the fitting are shown red in the residuals.
The observed and rest-frame wavelength are shown in the bottom and top of
the plot respectively.\label{fig_absspec}}
\end{figure}

We fitted all SDSS spectra using the {\sc nbursts} full spectrum fitting
technique \citep{CPSK07,CPSA07} and determined their radial velocities $v$,
stellar velocity dispersions $\sigma$, and parameterized star formation histories
represented by an instantaneous star burst (simple stellar populations, SSP)
or an exponentially declining star formation history (exp-SFH) assuming
that it started shortly after the Big Bang. We chose these two families of
stellar population models because: (i) SSP models are widely used in
extragalactic studies by different authors and we wanted our data to be
directly comparable to other sources; (ii) exponentially declining SFHs were
demonstrated to be a better representation of broadband SEDs of non-active
galaxies \citep{CZ12} than SSPs. We should, however, notice, that exp-SFH
models cannot adequately describe young stellar populations with mean ages 
$t<1.5$~Gyr (see discussion below).

The fitting procedure first convolves a grid of stellar population models
with the wavelength dependent spectral line spread function available for
every SDSS spectrum in the original data files, then runs a non-linear
Levenberg-Markquardt minimization by first choosing a model spectrum from
the grid by two-dimensional interpolation in the age--metallicity
($t$--[Fe/H]) space, then convolving it with a Gaussian-Hermite
representation of the line-of-sight velocity distribution (LOSVD) of stars
in a galaxy described by $v, \sigma, h3, h4$, and finally multiplying it by
a low order Legendre polynomial continuum (its parameters are determined
linearly in a separate loop) in order to absorb flux calibration
imperfections and possible internal extinction in a galaxy.  Hence, the
procedure returns values of $v, \sigma, h3, h4, t, $[Fe/H], and coefficients
of the multiplicative polynomial continuum.  Here we use a pure Gaussian
LOSVD shape with $h3=h4=0$.

The {\sc nbursts} algorithm is similar to the penalized pixel fitting
approach by \citet{CE04}.  It, however, has some important differences.  (i)
We use a linear fit of the low order multiplicative polynomial continuum
because its parameters are decoupled from galaxy kinematics and stellar
populations.  (ii) Instead of using a fixed grid of template spectra and
interpreting stellar populations using their relative weights in a linear
combination, we interpolate in a grid of models inside the minimization loop
in order to obtain the best-fitting stellar population parameters of each
starburst (or an exponentially declining model). As a result, for the
simplest case of a single component SSP model, we obtain the best-fitting
SSP-equivalent age and metallicity. These values are usually close to the
luminosity weighted ones, however, in cases of complex SFHs approximated by
an SSP there might be biases similar to those affecting Lick indices
\citep{ST07}. \citet{CPSA07,Chilingarian+08} demonstrated that SSP
equivalent ages and metallicity remain unbiased for galaxies with
super-solar $\alpha$-element abundances ([Mg/Fe]$>$0~dex) and when Balmer
line regions are masked in order to fit emission lines.

We excluded the spectral regions affected by bright atmosphere lines (O{\sc
i}, NaD, OH, etc.) and by the $A$ and $B$ telluric absorption bands from the
fitting procedure.  We also re-ran the fitting code excluding
$8\textup{--}14$~\AA-wide regions around locations of bright emission lines
for objects, where the reduced $\chi^2$ value of the fit exceeded the
threshold $\chi^2/DOF=$0.8, that was selected empirically from a sample of
galaxies without and with emission lines of different intensity
levels.\footnote{Published SDSS spectra are slightly oversampled in
wavelength, therefore, flux uncertainties in neighboring pixels are
correlated and, hence, the reduced $\chi^2$ for a spectrum well represented
by its model is less than 1 (around 0.6).}

We used three grids of stellar population models all computed with the
{\sc pegase.hr} evolutionary synthesis code \citep{LeBorgne+04}:
\begin{enumerate} 
\item SSP models based on the high resolution (R=10000)
ELODIE.3.1 empirical stellar library \citep{PS04,PSKLB07} covering the
wavelength range $3900 < \lambda < 6800$~\AA, the metallicity range $-2.5 <
$[Fe/H]$<0.5$~dex, and ages $20 < t < 20000$~Myr.
\item Models with 
exponentially declining SFH at a constant metallicity computed 
for the same metallicity and wavelength ranges as those for SSP
models, covering the range of exponential decay timescales $10 < \tau
< 20600$~Myr (the latter one effectively being a constant star formation rate
model) and starting epochs of star formation between $4.3$~Gyr and $13.8$~Gyr
of the age of the Universe corresponding to the redshift range $0<z<1.5$. We used the exponential
decay timescale $\tau$ in the same fashion as the SSP age in the minimization
procedure. For every galaxy, we first computed a grid of $\tau - $[Fe/H] models
with the star formation epoch corresponding to its redshift assuming that
a galaxy was formed at very high redshift, e.g. for $z=0.2$ with the light
travel time of 2.45~Gyr, we computed a grid of models for star
formation that started 11.27~Gyr ago assuming the standard WMAP9 cosmology \citep{Hinshaw+13}.
\item Intermediate resolution SSP models ($R=2300$) based on the MILES 
empirical stellar library \citep{SanchezBlazquez+06} 
covering the wavelength range $3600 < \lambda < 7400$~\AA, 
the metallicity range $-2.5 < $[Fe/H]$<0.7$~dex, and ages $20 < t < 20000$~Myr.
\end{enumerate}

We stress that the best-fitting stellar population ages $t>14000$~Myr (SSP) and 
exponential timescales $\tau<1000$~Myr (exp-SFH) should be considered as upper
and lower limits for the corresponding parameters.

In the public version of our catalog we provide two sets of stellar
population parameters for every galaxy: (1) SSP ages and metallicities
obtained from the spectrum fitting in the wavelength range in a galaxy
rest-frame $4500 < \lambda < 6795$~\AA\ using the MILES-{\sc pegase.hr}
models with the 5-th degree of the multiplicative polynomial continuum; (2)
{\sc pegase.hr} based exponentially declining SFR models in the wavelength
range $3915 < \lambda < 6795$~\AA\ with the 19-th degree continuum.  The
19-th degree corresponds to the emipirically determined optimal degree of
the multiplicative polynomial continuum for SDSS spectra when the $\chi^2$
value reaches a ``plateau'' as explained in \citet{Chilingarian+08}.  We
performed the SSP fitting with the MILES-{\sc pegase.hr} models in the
truncated wavelength range with a very low order polynomial continuum in
order to minimize the artifacts originating from imperfections in the SSP
model grid (see Section~4.2). In the publicly available Simple Spectrum Access
Service we provide the results of the MILES-{\sc pegase.hr} based SSP
fitting in the wavelength range $3600 < \lambda < 6790$~\AA\ in order to
enable the emission line analysis from the fitting residuals for all lines
including the [O{\sc ii}] 3727~\AA\ doublet.

\subsection{The spectral catalog: emission lines}

Our full spectral fitting procedure precisely matches the stellar continuum of each galaxy by the best-fitting stellar population model (see example in Fig.~\ref{fig_absspec}). 
Although the regions of all Balmer absorption lines are age sensitive, 
they contain at most 20\%\ of the age sensitive information from the entire
optical spectral range \citep{Chilingarian09}. \citet{CPSA07} have
demonstrated that masking the H$\beta$ and H$\gamma$ regions biases neither 
age nor metallicity determinations by the {\sc nbursts} procedure. Hence, we 
do not expect to introduce significant template mismatch by masking the 
regions of emission lines when fitting SDSS spectra.
Having subtracted the best-fitting model we obtain clean emission line spectra unaffected by stellar absorptions that is especially important for the Balmer lines.
The precision of our stellar continuum fitting allows us to recover faint emission lines at a few per~cent level of the continuum intensity, whereas very often such lines are not detected in the SDSS spectral pipeline results.
In Table~\ref{tbl_linelist} we provide the statistics of the emission line detection and strength in our sample.

In order to measure fluxes and equivalent widths (EW) of emission lines we applied two different approaches, namely Gaussian and non-parametric fitting of emission lines profiles.

In some galaxies, emission lines profiles cannot be described by a Gaussian.
This often becomes a case in galaxies with peculiar gas kinematics, e.g. multi-component bulk gas motions and outflows can produce complex asymmetric lines.
Also, this is crucially important for active galactic nuclei (AGN) with 
broad components in Balmer lines.
Approximation of such emission lines by a Gaussian profile results in biased estimates of flux and kinematic parameters.
We address this problem by employing a non-parametric fitting approach which allows us to recover arbitrary line profiles and measure their fluxes with higher precision.
At the same time, this method requires several lines with sufficiently high signal-to-noise ratio to be present in a spectrum, and may produce biased results when dealing with noisy data.
We, therefore, perform a ``classical'' Gaussian profile fitting too in order to allow for cross-comparison and validation of our line fitting results.

Both non-parametric and Gaussian fitting techniques take into account the SDSS
line-spread function computed individually for each spectrum by the standard SDSS pipeline and provided in FITS (Flexible Image Transport System) tables in the RCSED distribution.

\begin{figure*}
\includegraphics[clip,trim={3cm 0 3cm 0},width=\hsize]{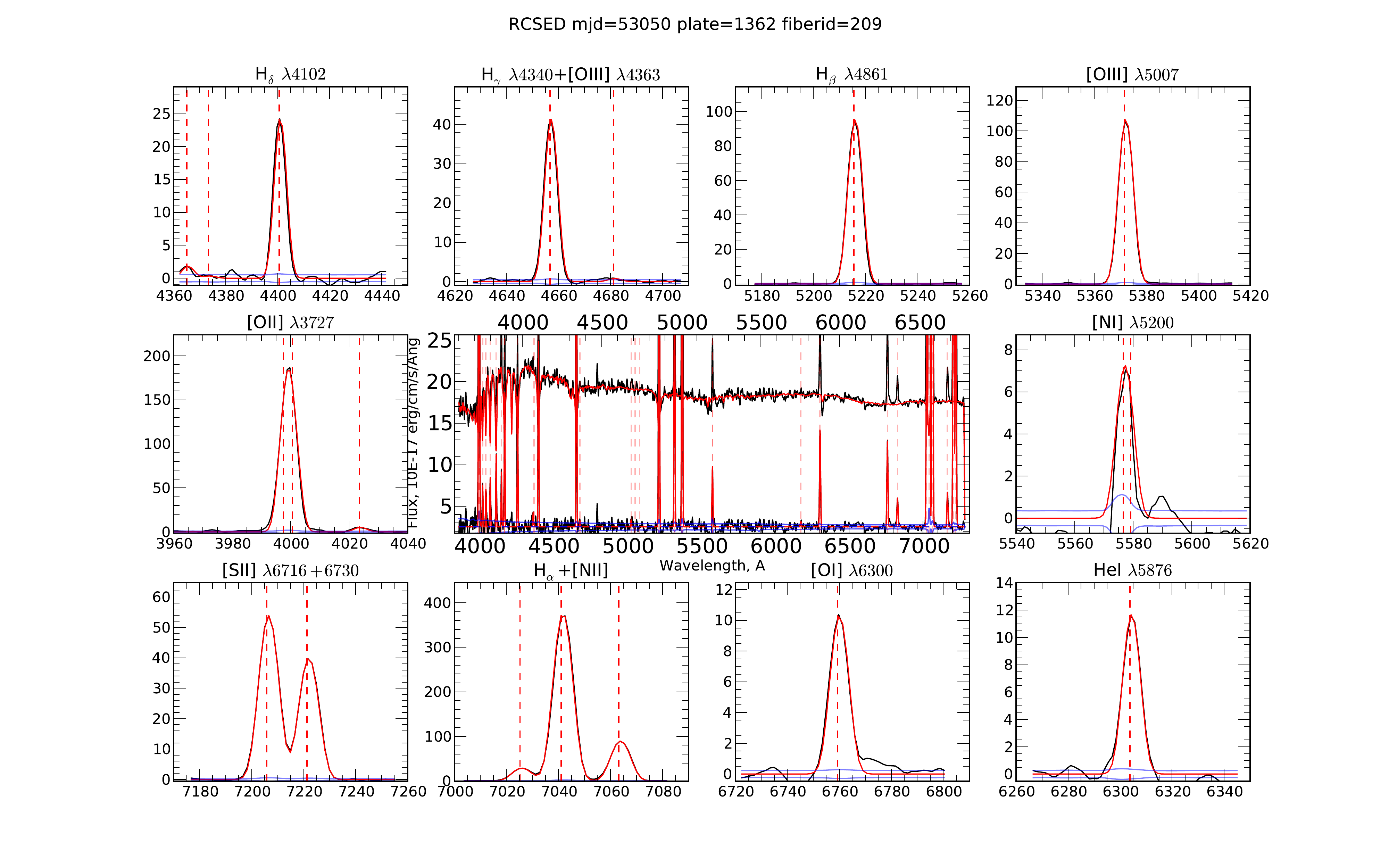}
\caption{Example of the {\sc nbursts} full spectrum fitting for an SDSS
spectrum of a late type galaxy together with the emission line fitting. 
Central panel is similar to Fig.~\ref{fig_absspec}, panels on the sides
demonstrate recovered profiles after the continuum subtraction of individual emission lines (black) and the 
best-fitting models (red). Blue lines show emission line flux uncertainties. Vertical red dashed lines represent the galaxy redshift in the SDSS database.
\label{fig_emspec}}
\end{figure*}

\subsubsection{Gaussian fitting}
\label{sec_gaussfit}

This approach consists of simultaneously fitting the entire set of emission
lines (see the line list in Table~\ref{tbl_linelist}) with Gaussians pre-convolved with the SDSS line-spread function. 
We allow two different sets of redshifts and intrinsic widths for recombination and forbidden lines. 
We estimate the kinematic parameters with the non-linear least-square minimization that is implemented by the MPFIT
package\footnote{\url{http://www.physics.wisc.edu/~craigm/idl/fitting.html}} \citep{Markwardt09}. 
The emission line fluxes are computed linearly for each
minimization iteration. 
When solving the linear problem, we constrain the line fluxes to be non-negative.
For this purpose we use the BVLS (bounded-variables least-squares) algorithm \citep{LH95} and its implementation by M.~Cappellari\footnote{\url{http://www-astro.physics.ox.ac.uk/~mxc/software/bvls.pro}}.

\begin{deluxetable*}{lclrrrrrrrrr}
\tablecolumns{12}
\tablecaption{Emission line detection statistics (the parametric Gaussian fit) at different signal-to-noise ratio (SNR). 
The ``Covered'' column reflects the number of objects with a corresponding
line in the wavelength coverage. The ``Wavelength'' column provides air
wavelengths. The ``Prefix'' column gives the prefix of column names for
the corresponding spectral line in the emission line FITS tables from the RCSED distribution.\label{tbl_linelist}}
\tablehead{	\colhead{Line} 				& 
			\colhead{Wavelength} 		&
			\colhead{Prefix} 		&
			\colhead{Covered} 			& 
			\multicolumn{2}{c}{$SNR>1$} & 
			\multicolumn{2}{c}{$SNR>3$} & 
			\multicolumn{2}{c}{$SNR>5$} & 
			\multicolumn{2}{c}{$SNR>10$}\\
			\colhead{} 					&
			\colhead{\AA} 				&
			\colhead{} 					&
			\colhead{N}					&
			\colhead{N}					&
			\colhead{fraction}			&
			\colhead{N}					&
			\colhead{fraction}			&		
			\colhead{N}					&
			\colhead{fraction}			&
			\colhead{N}					&
			\colhead{fraction}
}
\startdata

{[O\,{\sc ii}]       } &  3726.03 & f3727\_oii           &   780665 &   543374 &   69.6\%  &   354307 &   45.4\% &   225669 &   28.9\% &    92161 &  11.81\% \\
{[O\,{\sc ii}]       } &  3728.82 & f3730\_oii           &   782417 &   562707 &   71.9\%  &   387264 &   49.5\% &   257713 &   32.9\% &   110287 &  14.10\% \\
{H$\kappa$           } &  3750.15 & f3751\_h\_kappa      &   791720 &   233850 &   29.5\%  &    26219 &    3.3\% &     5584 &    0.7\% &      535 &   0.07\% \\
{H$\iota$            } &  3770.63 & f3772\_h\_iota       &   796081 &   180306 &   22.6\%  &    18912 &    2.4\% &     4524 &    0.6\% &      595 &   0.07\% \\
{H$\theta$           } &  3797.90 & f3799\_h\_theta      &   798386 &   229961 &   28.8\%  &    31590 &    4.0\% &     8466 &    1.1\% &     1345 &   0.17\% \\
{H$\eta$             } &  3835.38 & f3836\_h\_eta        &   798565 &   255213 &   32.0\%  &    52680 &    6.6\% &    16876 &    2.1\% &     2976 &   0.37\% \\
{[Ne\,{\sc iii}]     } &  3868.76 & f3870\_neiii         &   798635 &   246202 &   30.8\%  &    45733 &    5.7\% &    19500 &    2.4\% &     7579 &   0.95\% \\
{He\,{\sc i}         } &  3887.90 & f3889\_hei           &   798674 &   182764 &   22.9\%  &    37166 &    4.7\% &    10644 &    1.3\% &     1328 &   0.17\% \\
{H$\zeta$            } &  3889.07 & f3890\_h\_zeta      &   798675 &   223107 &   27.9\%  &    63940 &    8.0\% &    26166 &    3.3\% &     6072 &   0.76\% \\
{H$\epsilon$         } &  3970.08 & f3971\_h\_epsilon    &   798835 &   360759 &   45.2\%  &   156226 &   19.6\% &    78430 &    9.8\% &    21312 &   2.67\% \\
{[S\,{\sc ii}]       } &  4068.60 & f4070\_sii           &   799003 &   202235 &   25.3\%  &    14766 &    1.8\% &     2352 &    0.3\% &      149 &   0.02\% \\
{[S\,{\sc ii}]       } &  4076.35 & f4078\_sii           &   799010 &   149342 &   18.7\%  &     4755 &    0.6\% &      517 &    0.1\% &       55 &   0.01\% \\
{H$\delta$           } &  4101.73 & f4103\_h\_delta      &   799043 &   342858 &   42.9\%  &   172913 &   21.6\% &    96748 &   12.1\% &    32463 &   4.06\% \\
{H$\gamma$           } &  4340.46 & f4342\_h\_gamma      &   799276 &   419668 &   52.5\%  &   275775 &   34.5\% &   192540 &   24.1\% &    87637 &  10.96\% \\
{[O\,{\sc iii}]      } &  4363.21 & f4364\_oiii          &   799293 &   118667 &   14.8\%  &     8001 &    1.0\% &     2569 &    0.3\% &      787 &   0.10\% \\
{He\,{\sc ii}        } &  4685.76 & f4687\_heii          &   799381 &   109369 &   13.7\%  &     6779 &    0.8\% &     2204 &    0.3\% &      614 &   0.08\% \\
{[Ar\,{\sc iv}]      } &  4711.37 & f4713\_ariv          &   799381 &    79310 &    9.9\%  &     3477 &    0.4\% &      530 &    0.1\% &      110 &   0.01\% \\
{[Ar\,{\sc iv}]      } &  4740.17 & f4742\_ariv          &   799380 &   118077 &   14.8\%  &     7031 &    0.9\% &     1108 &    0.1\% &       85 &   0.01\% \\
{H$\beta$            } &  4861.36 & f4863\_h\_beta       &   799375 &   514321 &   64.3\%  &   381556 &   47.7\% &   317350 &   39.7\% &   214953 &  26.89\% \\
{[O\,{\sc iii}]      } &  4958.91 & f4960\_oiii          &   799372 &   449021 &   56.2\%  &   164285 &   20.6\% &    92442 &   11.6\% &    47287 &   5.92\% \\
{[O\,{\sc iii}]      } &  5006.84 & f5008\_oiii          &   799371 &   638852 &   79.9\%  &   404135 &   50.6\% &   244845 &   30.6\% &   119215 &  14.91\% \\
{[N\,{\sc i}]        } &  5197.90 & f5199\_ni            &   799360 &   144430 &   18.1\%  &     9742 &    1.2\% &     1345 &    0.2\% &      178 &   0.02\% \\
{[N\,{\sc i}]        } &  5200.25 & f5202\_ni            &   799360 &   184255 &   23.1\%  &    16318 &    2.0\% &     2676 &    0.3\% &      226 &   0.03\% \\
{[N\,{\sc ii}]       } &  5754.59 & f5756\_nii           &   799131 &   196670 &   24.6\%  &    12800 &    1.6\% &     2763 &    0.3\% &      966 &   0.12\% \\
{He\,{\sc i}         } &  5875.62 & f5877\_hei           &   798546 &   260312 &   32.6\%  &    81904 &   10.3\% &    38829 &    4.9\% &    12209 &   1.53\% \\
{[O\,{\sc i}]        } &  6300.30 & f6302\_oi            &   784763 &   439640 &   56.0\%  &   177144 &   22.6\% &    77850 &    9.9\% &    16856 &   2.15\% \\
{[O\,{\sc i}]        } &  6363.78 & f6366\_oi            &   780219 &   285886 &   36.6\%  &    40626 &    5.2\% &     9143 &    1.2\% &     1395 &   0.18\% \\
{[N\,{\sc ii}]       } &  6548.05 & f6550\_nii           &   764832 &   596254 &   78.0\%  &   422810 &   55.3\% &   289961 &   37.9\% &   133553 &  17.46\% \\
{H$\alpha$           } &  6562.79 & f6565\_h\_alpha      &   763451 &   614029 &   80.4\%  &   531966 &   69.7\% &   479842 &   62.9\% &   395722 &  51.83\% \\
{[N\,{\sc ii}]       } &  6583.45 & f6585\_nii           &   761376 &   641883 &   84.3\%  &   553212 &   72.7\% &   479386 &   63.0\% &   334901 &  43.99\% \\
{He\,{\sc i}         } &  6678.15 & f6679\_hei           &   750612 &   178330 &   23.8\%  &    21069 &    2.8\% &     6321 &    0.8\% &     1408 &   0.19\% \\
{[S\,{\sc ii}]       } &  6716.43 & f6718\_sii           &   745687 &   571758 &   76.7\%  &   423064 &   56.7\% &   320126 &   42.9\% &   186973 &  25.07\% \\
{[S\,{\sc ii}]       } &  6730.81 & f6733\_sii           &   743742 &   554071 &   74.5\%  &   374143 &   50.3\% &   263155 &   35.4\% &   135572 &  18.23\% \\
\enddata
\end{deluxetable*}

\subsubsection{Non-parametric emission line fitting}

\begin{figure}
\includegraphics[width=\hsize]{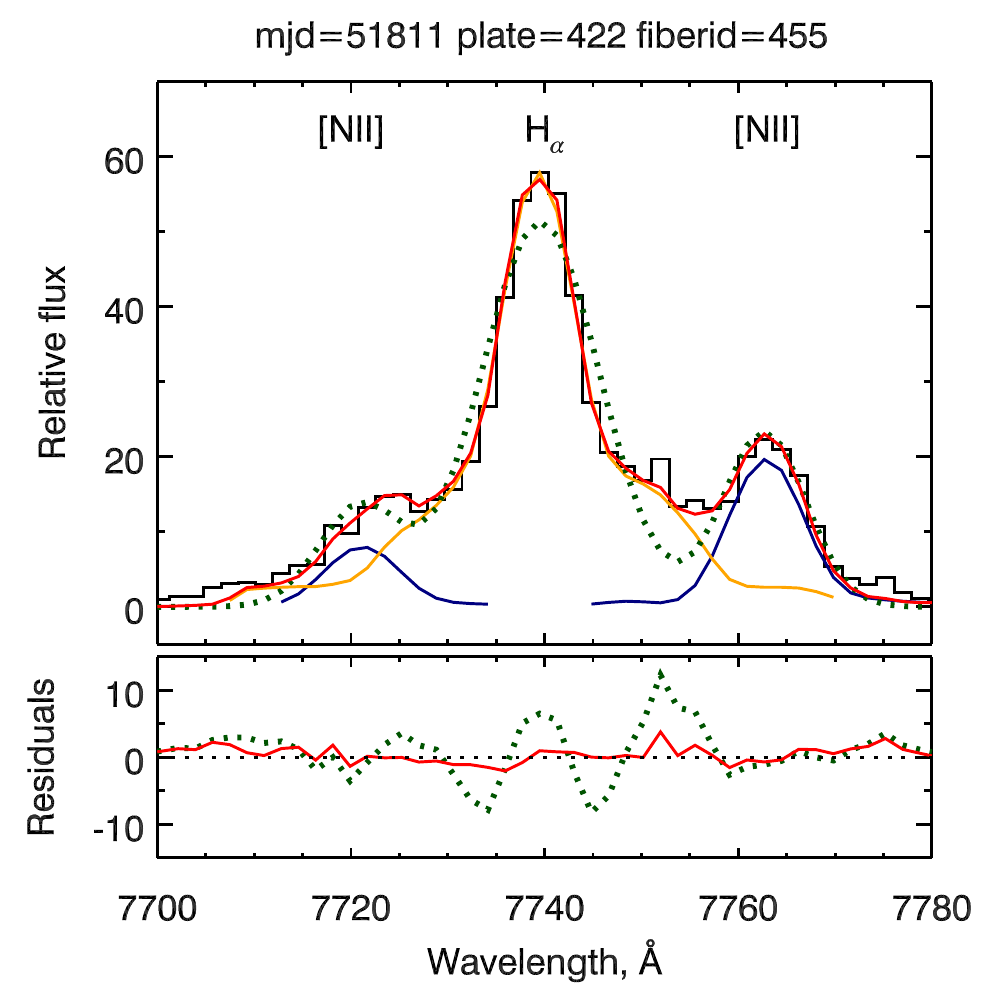}
\caption{An example of the complex emission line profile of a Seyfert galaxy,
and results of its fitting with two different techniques. Black stepped line in
the upper panel shows the observed spectrum of H$\alpha$ and N{\sc ii}
lines in relative flux units, green dotted line is a Gaussian fit result, red
solid line is a non-parametric fitting result. Individual H$\alpha$ and 
N{\sc ii} profiles recovered by the non-parametric fitting are shown in
orange and blue respectively. Lower panel shows fitting residuals. In
the case of complex asymmetric emission lines profiles non-parametric fitting
method is clearly preferred over Gaussian one.
\label{fig_emis_lines}}
\end{figure}

Our non-parametric emission line fitting method includes two main steps which we repeat several times until the convergence is achieved.
First, we derive discretely sampled emission line profiles, i.e. line-of-sight velocity distributions (LOSVDs) of ionized gas.
During the second step, we estimate emission lines fluxes. Because allowed
and forbidden transitions often originate from different regions of a galaxy
having very different physical properties (i.e.  density, temperature,
mechanism of excitation), however, all emission lines of each type (allowed
and forbidden) have similar shapes, our procedure recovers two
different non-parametric profiles, one for each type.

The LOSVD derivation is organized as follows.
We note that convolution of any logarithmically rebinned observed spectrum $S_{obs}$ of $m$ elements with LOSVD $\mathcal{L}$ having $n$ elements can be expressed as a linear matrix equation $A \ast \mathcal{L} = S_{obs}$, where $A$ is a $m\times n$ matrix of template spectra having lengths of $m$ pixels each.
Every template spectrum from the $i$-th row in the matrix $A$ is shifted by the velocity which represents the $i$-th position within the LOSVD vector.
Here a template spectrum is a synthetic spectrum made of a set of flux normalized Gaussians with LSF widths representing emission lines detected in the observed spectrum.
Such approach allows us to take into account the SDSS instrumental resolution instead of a set of Dirac $\delta$-functions.
The continuum level of a template spectrum is set to zero.

Thus, we end up with a linear inverse problem whose solution $\mathcal{L}$ can be derived by a least square technique. 
The emission line profiles obviously cannot be negative and, therefore, we use the BVLS algorithm mentioned above.

Once the LOSVD has been derived, we compute emission line fluxes by solving a similar linear problem to that described in Section~\ref{sec_gaussfit}.
This finishes the first iteration.

At the same time, the LOSVD derivation step requires the knowledge of emission line fluxes in order to construct better template spectra.
During the first iteration when they are unknown, we set all fluxes to unity and the all template spectra hence are made of equally normalized Gaussians.
Typically, 3 iterations of this procedure is enough to reach the convergence.

A linear inversion is an ill-conditioned problem and is sensitive
to noise in the data. In order to improve the profile
reconstruction quality, we exploit a regularization
technique, which minimizes the squared third derivative of the recovered
line profile. This approach, however, causes artefacts in sharp narrow line
profiles. Therefore, we apply the regularization only in the wings of
emission lines where flux levels are generally low and, consequently, noise
is higher. The regularization technique yields the dramatic improvement of
recovered Balmer line profiles for faint AGNs. In the catalog we provide
measurements of non-parametric emission lines with and without
regularization.

The comparison between the parametric (Gaussian) and non-parametric fitting results for a complex emission line profile in a Seyfert galaxy is presented at Fig.~\ref{fig_emis_lines}.
A Gaussian approximation for such lines is often inadequate and causes serious biases in the kinematics that can reach few hundred km s$^{-1}$.

We ran Monte-Carlo simulations for a random sample of 2,000 objects with
emission lines of different intensity levels in order to estimate realistic flux
uncertainties obtained with the non-parametric fitting technique. They
turned out to be consistent with statistical uncertainties of Gaussian
emission line fluxes for most objects and up-to a factor of 2 lower for AGNs
with broad line components. The RCSED database will be updated with
Monte-Carlo based uncertainties as we compute them: this procedure is very
computationally intensive and will take a couple of months to complete.

\subsubsection{Gas phase metallicities}

We used our emission line flux measurements in order to estimate the gas
phase metallicities for galaxies where emissions originate from the star
formation induced excitation.  We exploited two different techniques to
measure the metallicity, (i) a new calibration by \citet{DKSN16} and (ii)
the IZI Bayesian technique \citep{BKVD15} using a grid of models
($\kappa=\infty$) with $\kappa$-distributed electron energies
\citep{Dopita+13}.  We selected star formation
dominated and ``transition type'' galaxies using the standard BPT
\citep{BPT81} diagram that exploits hydrogen, nitrogen, and oxygen emission
lines with the criteria defined in \citet{Kauffmann+03}.

The \citet{DKSN16} calibration uses only the H$\alpha$, [N{\sc ii}], and [S{\sc ii}]
emission lines, all located in a very narrow spectral interval and is, therefore,
virtually insensitive to the internal extinction within an observed galaxy. 
This calibration is presented in a form of a simple formula which makes it
very easy to use.  However, a disadvantage of this approach in 
application to our dataset is that at redshifts $z>0.1$ the emission lines
used for the metallicity determination shift to the spectral region
dominated by telluric absorption and airglow emission lines (mostly, OH)
which can seriously affect the quality of the emission line flux estimates. 
Another natural limitation of this approach originates from the SDSS spectral
wavelength range ($\lambda<9200$~\AA) that corresponds to the upper redshift
limit $z=0.36$ when the forbidden sulfur line [S{\sc ii}] 6730~\AA\ shifts out of the wavelength
range.  Besides, the calibration critically depends on the [N/O] relation
and is therefore sensitive to possible galaxy to galaxy [N/H] abundance
variations. In the catalog we included the metallicity estimates obtained
using the \citet{DKSN16} calibration for Gaussian
emission line analysis ({\tt rcsed\_gasmet} table). We computed the uncertainties of the gas phase
metallicities by propagating the statistical flux errors through the
calculations according to the formula in \citet{DKSN16}.

The IZI technique \citep{BKVD15} takes advantage of all available emission
line measurements and, hence, is more robust and can in principle be used
for the entire sample of SDSS galaxies.  The algorithm is implemented in an
{\sc idl} software package distributed by the authors along with 17 grids of
photoionization models.  However, this technique relies on the external dust
attenuation correction which must be applied to emission lines fluxes prior
to fitting.  It also requires (similar to the \citet{DKSN16} approach) a
pre-selection of star forming galaxies.  We estimated the internal dust
attenuation using the typical value of the Balmer decrement
H$\alpha$/H$\beta = 2.83$ \citep{GBW12} and corrected all emission line
fluxes accordingly.  In galaxies where the observed H$\alpha$/H$\beta$ ratio
fell below that value, we assumed the extinction to be zero.  Finally, the
fluxes were supplied to the IZI software package with the \citet{Dopita+13}
model grid and the resulting [O/H] and ionizing parameter values for 
Gaussian emission line fluxes were included in the
gas phase metallicity table {\tt rcsed\_gasmet} of the catalog.

\section{Photometric properties of the sample}

\begin{figure}
\includegraphics[width=\hsize]{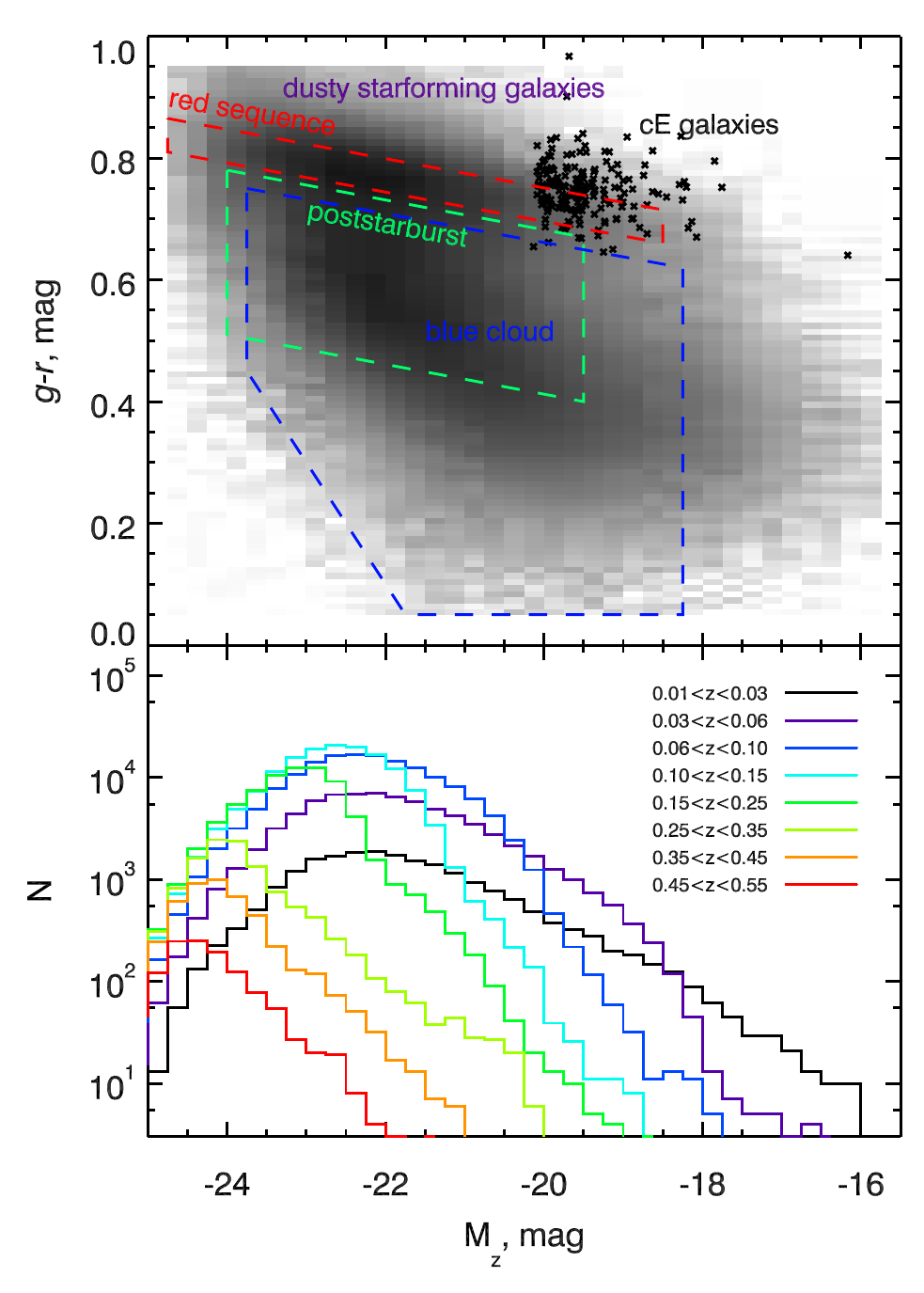}
\caption{(Top) Optical color--magnitude diagram for extinction- and
$k$-corrected Petrosian magnitudes of all galaxies in our sample. (Bottom)
Redshift distributions of galaxies in corresponding bins on absolute
magnitude.\label{fig_red_sequence}}
\end{figure}

\subsection{Completeness at different redshifts}

Because our catalog uses the SDSS DR7 spectroscopic galaxy sample as its
master list, and the legacy SDSS spectroscopic survey was magnitude limited
with the $r=17.77$~mag limit in a 3~arcsec aperture, we sample different
parts of the galaxy luminosity function with the redshift dependent
completeness. Also, there is an important fiber
collision effect, that is when two fibers in the SDSS multi-object
spectrograph cannot be put too close to each other: because of this, there
is a systematic undersampling of dense clusters and groups of galaxies.

In Fig.~\ref{fig_red_sequence} (top panel) we present a two-dimensional
distribution of our galaxies in the $(M_z, g-r)$ color--magnitude space. 
We identify the regions traditionally referred to as ``the red sequence''
and ``the blue cloud'' as well as the locus of typical post-starburst (E+A)
galaxies.  The density in the plot corresponds to the object number density
in our catalog at a given position of the parameter space.  We also show
by small crosses the tidally stripped systems, compact elliptical galaxies,
from the sample of \citet{CZ15} which reside systematically above the red
sequence. One can see the bimodality of the galaxy distribution by color
for intermediate luminosity and dwarf galaxies ($M_z>-20.5$~mag) while the
transition is rather smooth for more luminous systems. 

Even though dwarf galaxies are more numerous in the Universe than giants
because of the raising low end of the galaxy luminosity function
\citep{Schechter76,Blanton+03a}, we see the apparent decrease of the
histogram density at fainter magnitudes.  In the bottom panel of
Fig.~\ref{fig_red_sequence} we demonstrate the breakdown by redshift for the
luminosity distribution of galaxies contributing to the histogram in the top
panel.  The high luminosity end decline is due to the intrinsic shape of the
luminosity function, while the low luminosity tail drops because of the SDSS
completeness and target selection biased against very extended (and
therefore nearby) galaxies.  We clearly see how the magnitude limit
constraint of SDSS causes the drop in the number of galaxies further and
further up the luminosity function as we move to higher redshifts. 
Fig.~\ref{fig_red_sequence} confirms that we start probing the dwarf galaxy
regime ($M_z>-19.8$~mag) at $z<0.06$, however the selection effects have to
be seriously considered for any type of a statistical study.

\subsection{Red sequence in different bands}

For practical reasons such as selection of candidate early-type members in
galaxy clusters using photometric data, it is important to know the shape of
the red sequence in different photometric bands. Here we provide the best
fitting second degree polynomial approximations of the red sequence shape
for a set of galaxy colors spanning optical and NIR bands.

First we created a sample of red sequence galaxies by the following
criteria: (1) We selected all objects at redshifts $z<0.27$; (2) we applied
a color cut on $NUV-r$ colors by selecting all objects on $(M_r,NUV-r)$
plane that resided above the straight line passing through
$p_0=(-16.0,3.5)$~mag and $p_1=(-24.0,5.0)$~mag; (3) we applied a color cut
on $g-r$ colors by selecting all objects on $(M_r,g-r)$ plane that resided
above the straight line passing through $q_0=(-16.0,0.5)$~mag and
$q_1=(-24.0,0.75)$~mag and also satisfying the criterion $(g-r)<0.95$~mag.

Then, in order to account for two orders of magnitude variations of galaxy
density along the red sequence, for every combination of colors and magnitudes
(e.g.  $g-r$, $M_r$): (1) We selected measurements having statistical
uncertainties $<0.1$~mag in both bands; (2) binned the distribution on
luminosity using 0.5~mag wide bins and computed median color values and
outlier resistant standard deviations in every bin; (3) fitted a second
degree polynomial into median values only in those bins that contained more
than 15 objects.  For convenience and because mean absolute $AB$ magnitudes
of galaxies in our sample stay mostly within the range $-25<M<-15$~mag in
all optical red ($riz$) and NIR filters, we added 20.0~mag to all absolute
magnitudes prior to fitting.

\begin{flalign}
(u-r) &=  +2.51 -0.065  \cdot M_{20r} -0.005  \cdot M_{20r}^2;\, \sigma= 0.16 \nonumber \\
(u-i) &=  +2.90 -0.069  \cdot M_{20i} -0.007  \cdot M_{20i}^2;\, \sigma= 0.17 \nonumber \\
(u-z) &=  +3.15 -0.050  \cdot M_{20z} -0.014  \cdot M_{20z}^2;\, \sigma= 0.19 \nonumber \\
(g-r) &=  +0.75 -0.026  \cdot M_{20r} -0.001  \cdot M_{20r}^2;\, \sigma= 0.045 \nonumber \\
(g-i) &=  +1.12 -0.038  \cdot M_{20i} -0.003  \cdot M_{20i}^2;\, \sigma= 0.074 \nonumber \\
(g-z) &=  +1.39 -0.044  \cdot M_{20z} -0.009  \cdot M_{20z}^2;\, \sigma= 0.10 \nonumber \\
(g-Y) &=  +1.91 -0.067  \cdot M_{20Y} -0.018  \cdot M_{20Y}^2;\, \sigma= 0.14 \nonumber \\
(g-J) &=  +2.01 -0.073  \cdot M_{20J} -0.016  \cdot M_{20J}^2;\, \sigma= 0.18 \nonumber \\
(g-H) &=  +2.30 -0.094  \cdot M_{20H} -0.015  \cdot M_{20H}^2;\, \sigma= 0.19 \nonumber \\
(g-K) &=  +2.00 -0.108  \cdot M_{20K} -0.018  \cdot M_{20K}^2;\, \sigma= 0.22 \nonumber \\
&M_{20\mathrm{col}} = M_{\mathrm{col}} + 20.0 \mathrm{mag}
\label{red_seq_eq}
\end{flalign}

Eqs.~\ref{red_seq_eq} provide the best-fitting polynomials for the red
sequence shape in 10 photometric bands.  We consider the mean standard
deviation value from all bins used in the fitting procedure as the ``red
sequence width'' ($\sigma$ in Eqs.~\ref{red_seq_eq}) and stress that the
actual fitting residuals for median values are usually an order of magnitude
smaller.

We notice that in the most widely used parameter spaces, $(M_r, g-r)$,
$(M_r, u-r)$, $(M_i, g-i)$, and $(M_z, g-z)$, the red sequence does not show
any substantial curvature which is indicated by negligible 2nd order
polynomial terms.  This suggests that there is no ``red sequence saturation''
at the bright end.

\subsection{Color transformations for galaxies of different morphologies
and luminosities}

\begin{table*}
\caption{Median rest-frame colors of galaxies of different morphological
types and luminosities in $AB$ magnitudes.  For every color (left column)
there are three groups corresponding to giant (1st group), intermediate
luminosity (2nd group), and dwarf (3rd group) galaxies of 6 values for 6
Hubble types.  Standard deviation values for each median color are presented in the
adjacent table rows.\label{tab_gal_col}}
\bgroup
\begin{tabular}{l|llllll|llllll|llllll}
\hline
 & \multicolumn{6}{c}{$-24.0 \le M_r < -22.0$~mag} & \multicolumn{6}{c}{$-22.0 \le M_r < -19.0$~mag} &\multicolumn{6}{c}{$-19.0 \le M_r < -16.0$~mag} \\
 & Sdm & Sc & Sb & Sa & S0 & E~~~~~~ & Sdm & Sc & Sb & Sa & S0 & E~~~~~~ & Sdm & Sc & Sb & Sa & S0 & E \\
\hline
$FUV$-$r$  &  1.72 &  2.76 &  3.46 &  4.33 &  5.58 &  6.86 &  1.60 &  2.48 &  3.22 &  4.15 &  5.22 &  6.72 &  1.44 &  2.18 &  2.87 &  3.81 &  4.91 &  6.78\\
stdev & \scriptsize{0.33} & \scriptsize{0.33} & \scriptsize{0.40} & \scriptsize{0.50} & \scriptsize{0.78} & \scriptsize{0.67} & \scriptsize{0.31} & \scriptsize{0.32} & \scriptsize{0.39}
& \scriptsize{0.48} & \scriptsize{0.71} & \scriptsize{0.78} & \scriptsize{0.26} & \scriptsize{0.31} & \scriptsize{0.36} & \scriptsize{0.46} & \scriptsize{0.75} & \scriptsize{0.91} \\
$NUV$-$r$  &  1.20 &  2.20 &  2.89 &  3.73 &  4.84 &  5.64 &  1.17 &  2.02 &  2.69 &  3.57 &  4.57 &  5.44 &  1.15 &  1.80 &  2.41 &  3.29 &  4.18 &  4.98\\
stdev & \scriptsize{0.26} & \scriptsize{0.20} & \scriptsize{0.27} & \scriptsize{0.29} & \scriptsize{0.30} & \scriptsize{0.25} & \scriptsize{0.22} & \scriptsize{0.24} & \scriptsize{0.27}
& \scriptsize{0.28} & \scriptsize{0.29} & \scriptsize{0.27} & \scriptsize{0.18} & \scriptsize{0.23} & \scriptsize{0.24} & \scriptsize{0.25} & \scriptsize{0.26} & \scriptsize{0.25} \\
$u$-$r$  &  0.99 &  1.48 &  1.80 &  2.14 &  2.44 &  2.56 &  0.95 &  1.34 &  1.68 &  2.05 &  2.32 &  2.44 &  0.88 &  1.19 &  1.44 &  1.76 &  2.02 &  2.14\\
stdev & \scriptsize{0.15} & \scriptsize{0.19} & \scriptsize{0.21} & \scriptsize{0.21} & \scriptsize{0.23} & \scriptsize{0.19} & \scriptsize{0.19} & \scriptsize{0.20} & \scriptsize{0.20}
& \scriptsize{0.21} & \scriptsize{0.21} & \scriptsize{0.17} & \scriptsize{0.26} & \scriptsize{0.20} & \scriptsize{0.19} & \scriptsize{0.22} & \scriptsize{0.24} & \scriptsize{0.21} \\
$g$-$r$  &  0.26 &  0.49 &  0.60 &  0.70 &  0.78 &  0.80 &  0.24 &  0.40 &  0.53 &  0.65 &  0.73 &  0.76 &  0.22 &  0.34 &  0.45 &  0.57 &  0.66 &  0.69\\
stdev & \scriptsize{0.15} & \scriptsize{0.06} & \scriptsize{0.05} & \scriptsize{0.05} & \scriptsize{0.05} & \scriptsize{0.03} & \scriptsize{0.09} & \scriptsize{0.07} & \scriptsize{0.06}
& \scriptsize{0.05} & \scriptsize{0.05} & \scriptsize{0.04} & \scriptsize{0.10} & \scriptsize{0.06} & \scriptsize{0.06} & \scriptsize{0.06} & \scriptsize{0.06} & \scriptsize{0.04} \\
$g$-$i$  &  0.42 &  0.75 &  0.92 &  1.06 &  1.15 &  1.17 &  0.33 &  0.58 &  0.81 &  1.00 &  1.10 &  1.12 &  0.26 &  0.45 &  0.64 &  0.85 &  0.98 &  1.02\\
stdev & \scriptsize{0.27} & \scriptsize{0.11} & \scriptsize{0.09} & \scriptsize{0.07} & \scriptsize{0.07} & \scriptsize{0.05} & \scriptsize{0.16} & \scriptsize{0.12} & \scriptsize{0.11}
& \scriptsize{0.08} & \scriptsize{0.08} & \scriptsize{0.06} & \scriptsize{0.16} & \scriptsize{0.11} & \scriptsize{0.10} & \scriptsize{0.11} & \scriptsize{0.10} & \scriptsize{0.07} \\
$g$-$z$  &  0.66 &  0.97 &  1.16 &  1.32 &  1.42 &  1.44 &  0.41 &  0.72 &  1.02 &  1.25 &  1.37 &  1.39 &  0.33 &  0.55 &  0.79 &  1.05 &  1.18 &  1.24\\
stdev & \scriptsize{0.32} & \scriptsize{0.18} & \scriptsize{0.14} & \scriptsize{0.11} & \scriptsize{0.10} & \scriptsize{0.07} & \scriptsize{0.21} & \scriptsize{0.18} & \scriptsize{0.16}
& \scriptsize{0.12} & \scriptsize{0.11} & \scriptsize{0.09} & \scriptsize{0.21} & \scriptsize{0.18} & \scriptsize{0.16} & \scriptsize{0.16} & \scriptsize{0.15} & \scriptsize{0.10} \\
$g$-$Y$  &  1.27 &  1.50 &  1.70 &  1.87 &  1.97 &  1.98 &  0.78 &  1.12 &  1.50 &  1.79 &  1.91 &  1.92 &  0.59 &  0.82 &  1.12 &  1.46 &  1.58 &  1.68\\
stdev & \scriptsize{0.45} & \scriptsize{0.19} & \scriptsize{0.17} & \scriptsize{0.13} & \scriptsize{0.12} & \scriptsize{0.09} & \scriptsize{0.31} & \scriptsize{0.27} & \scriptsize{0.22}
& \scriptsize{0.16} & \scriptsize{0.14} & \scriptsize{0.12} & \scriptsize{0.33} & \scriptsize{0.28} & \scriptsize{0.26} & \scriptsize{0.26} & \scriptsize{0.23} & \scriptsize{0.15} \\
$g$-$J$  &  1.37 &  1.56 &  1.77 &  1.96 &  2.05 &  2.07 &  0.83 &  1.15 &  1.58 &  1.90 &  2.01 &  2.03 &  0.53 &  0.77 &  1.11 &  1.51 &  1.66 &  1.75\\
stdev & \scriptsize{0.46} & \scriptsize{0.23} & \scriptsize{0.21} & \scriptsize{0.17} & \scriptsize{0.16} & \scriptsize{0.13} & \scriptsize{0.38} & \scriptsize{0.35} & \scriptsize{0.27}
& \scriptsize{0.19} & \scriptsize{0.18} & \scriptsize{0.15} & \scriptsize{0.49} & \scriptsize{0.46} & \scriptsize{0.39} & \scriptsize{0.36} & \scriptsize{0.32} & \scriptsize{0.22} \\
$g$-$H$  &  1.63 &  1.87 &  2.10 &  2.30 &  2.39 &  2.40 &  1.05 &  1.42 &  1.89 &  2.22 &  2.33 &  2.33 &  0.77 &  1.02 &  1.37 &  1.80 &  1.89 &  1.99\\
stdev & \scriptsize{0.54} & \scriptsize{0.24} & \scriptsize{0.21} & \scriptsize{0.18} & \scriptsize{0.17} & \scriptsize{0.14} & \scriptsize{0.39} & \scriptsize{0.35} & \scriptsize{0.27}
& \scriptsize{0.21} & \scriptsize{0.20} & \scriptsize{0.16} & \scriptsize{0.47} & \scriptsize{0.38} & \scriptsize{0.34} & \scriptsize{0.36} & \scriptsize{0.32} & \scriptsize{0.20} \\
$g$-$K$  &  1.54 &  1.61 &  1.83 &  2.03 &  2.11 &  2.11 &  0.80 &  1.12 &  1.60 &  1.95 &  2.04 &  2.03 &  0.34 &  0.62 &  1.00 &  1.44 &  1.55 &  1.63\\
stdev & \scriptsize{0.52} & \scriptsize{0.25} & \scriptsize{0.23} & \scriptsize{0.19} & \scriptsize{0.18} & \scriptsize{0.15} & \scriptsize{0.45} & \scriptsize{0.40} & \scriptsize{0.30}
& \scriptsize{0.24} & \scriptsize{0.23} & \scriptsize{0.18} & \scriptsize{0.56} & \scriptsize{0.49} & \scriptsize{0.42} & \scriptsize{0.44} & \scriptsize{0.36} & \scriptsize{0.24} \\
\hline
\end{tabular}
\egroup
\end{table*}

\citet{CZ12} demonstrated that the Hubble morphological classification
derived by a human eye \citep{Fukugita+07} correlates very well with the
total $NUV-r$ color of a galaxy.  With a computed dispersion of $0.8 t$,
where $t$ is the Hubble type, it corresponds to the subjective precision of
such a classification.  Here we use this relation in order to derive median
values of galaxy colors across the Hubble sequence for three galaxy luminosity classes defined
on a basis of their $r$-band luminosities.

We dissect the $(M_r, NUV-r)$ color--magnitude plane into 18 quadrangular
regions by assuming that the morphological type for giant galaxies
($M_r=-24$~mag) can be estimated by linearly varying the $(NUV-r)$ color
from $+0.5$ to $+6.5$~mag with a step of 1~mag corresponding to one Hubble
type from \emph{Sd} to \emph{E}.  At the same time, we assume that in the
dwarf regime $(M_r=-16$~mag) the step reduces to 0.75~mag per Hubble type
that corresponds to the observed reduction of the $(NUV-r)$ color range. 
We choose 3 luminosity bins, $-24.0 \le M_r< -22.0$~mag, $-22.0 \le M_r<
-19.0$~mag, and $-19.0 \le M_r <-16.0$~mag, which represent giant,
intermediate luminosity, and dwarf galaxies. Then, in every region we
compute the median value of a desired color, and the standard deviation of
the distribution.

We present our results in Table~\ref{tab_gal_col}. They expand and update the
widely used color transformations from \citet{FSI95} by using a very rich dataset properly
corrected for systematic effects and using modern prescriptions for
$k$-corrections.  We extend their results at $z=0$ (see table~3 in
\citealp{FSI95}) to near-UV and NIR colors and also towards intermediate
and low luminosity galaxies.  The direct comparison of our values with those
of \citet{FSI95} reveals a good agreement of optical colors except (a) S0
galaxies which are systematically redder in our case and stay really close
to the ellipticals; (b) the $u-g$ color of ellipticals that is some 0.25~mag
bluer in our case.  We assign the latter systematics to our improved
$k$-correction prescriptions for the $u$ band photometry and generally
higher quality of the $u$ band SDSS photometric data compared to the dataset
used in \citet{FSI95}.  On the other hand, we attribute redder colors of
lenticular galaxies in our data to the specificities of the synthetic color
estimation technique used in \citet{FSI95} that underestimated colors of 2 of
4 their lenticular galaxies by 0.1--0.15~mag (see their table~1).

\section{Spectroscopic properties of the sample}

\subsection{Stellar kinematics of galaxies}

In comparison to original SDSS measurements of stellar kinematics based
on cross-correlation with a limited set of template spectra, our approach
yields significantly smaller template mismatch between models and observed
spectra for non-active galaxies.  We, therefore, achieve on average
30\% lower statistical uncertainties of radial velocity and velocity
dispersion measurements.  Moreover, there is a known degeneracy between
stellar metallicity and velocity dispersion estimates when
using pixel space fitting techniques \citep{CPSA07}, because an underestimated metallicity
(i.e.  using a metal poor template for a metal rich galaxy) can be
compensated by a lower velocity dispersion that would smear that template
spectrum to a lesser degree.  Therefore, by using a grid of stellar population models
ranging from low ([Fe/H]$=-2.0$~dex) to high ([Fe/H]$=+0.7$~dex)
metallicities and covering the whole range of ages, we reduce the systematic
errors of velocity dispersion measurements, especially in the most
metal-rich regime including massive elliptical and lenticular galaxies.  On
the other hand, we accurately take into account the spectral line spread
function of the SDSS spectrograph that allows us to measure velocity
dispersions down to 50~km~s$^{-1}$, thus going far into the dwarf galaxy
regime \citep{Chilingarian09}.

\begin{figure}
\includegraphics[width=\hsize]{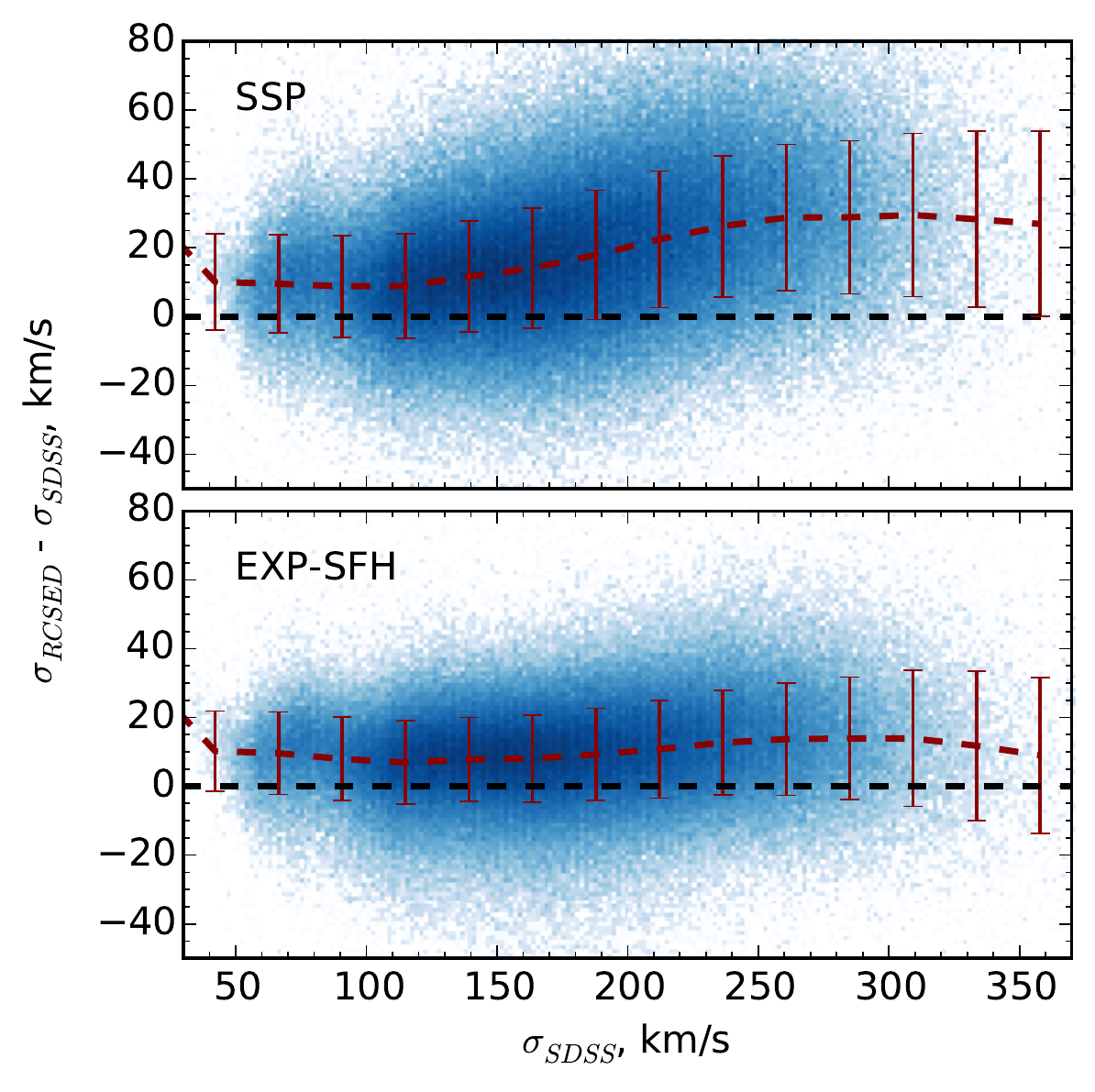}
\caption{Comparison of RCSED stellar velocity dispersion measurements
with those published by the SDSS DR7. Top and bottom panel correspond to the
two sets of stellar population models, SSP and exponentially declining
SFHs correspondingly. \label{figsigsig}}
\end{figure}

As it was already pointed out by \citet{Fabricant+13}, stellar velocity
dispersion measurements in the SDSS DR7 catalog are systematically
underestimated for luminous elliptical galaxies compared to the values
obtained by the full spectrum fitting, that is likely caused by
the template mismatch and degeneracy with metallicity mentioned above. Here
we observe a very similar trend: our SSP velocity dispersion measurements for
massive ellipticals ($\sigma \gtrsim 250$~km~s$^{-1}$) are up-to
30~km~s$^{-1}$ higher than those reported in the SDSS DR7 catalog, and this difference
goes down to 7--10~km~s$^{-1}$ for low luminosity galaxies ($\sigma \sim 100$~km~s$^{-1}$).
In Fig.~\ref{figsigsig} (top panel) we present the comparison made for our entire
sample for 361,421 galaxies with velocity dispersion uncertainties better than
7\% of the value (i.e. $\Delta \sigma = $7~km~s$^{-1}$ for $\sigma =
100$~km~s$^{-1}$). The velocity dispersions estimated from the fitting of
exponentially declining SFH models computed with {\sc pegase.hr}, are a
little bit closer to the values in SDSS DR7, however, the general trend
looks similar (Fig.~\ref{figsigsig}, bottom panel).

\begin{figure*}
\includegraphics[width=\hsize]{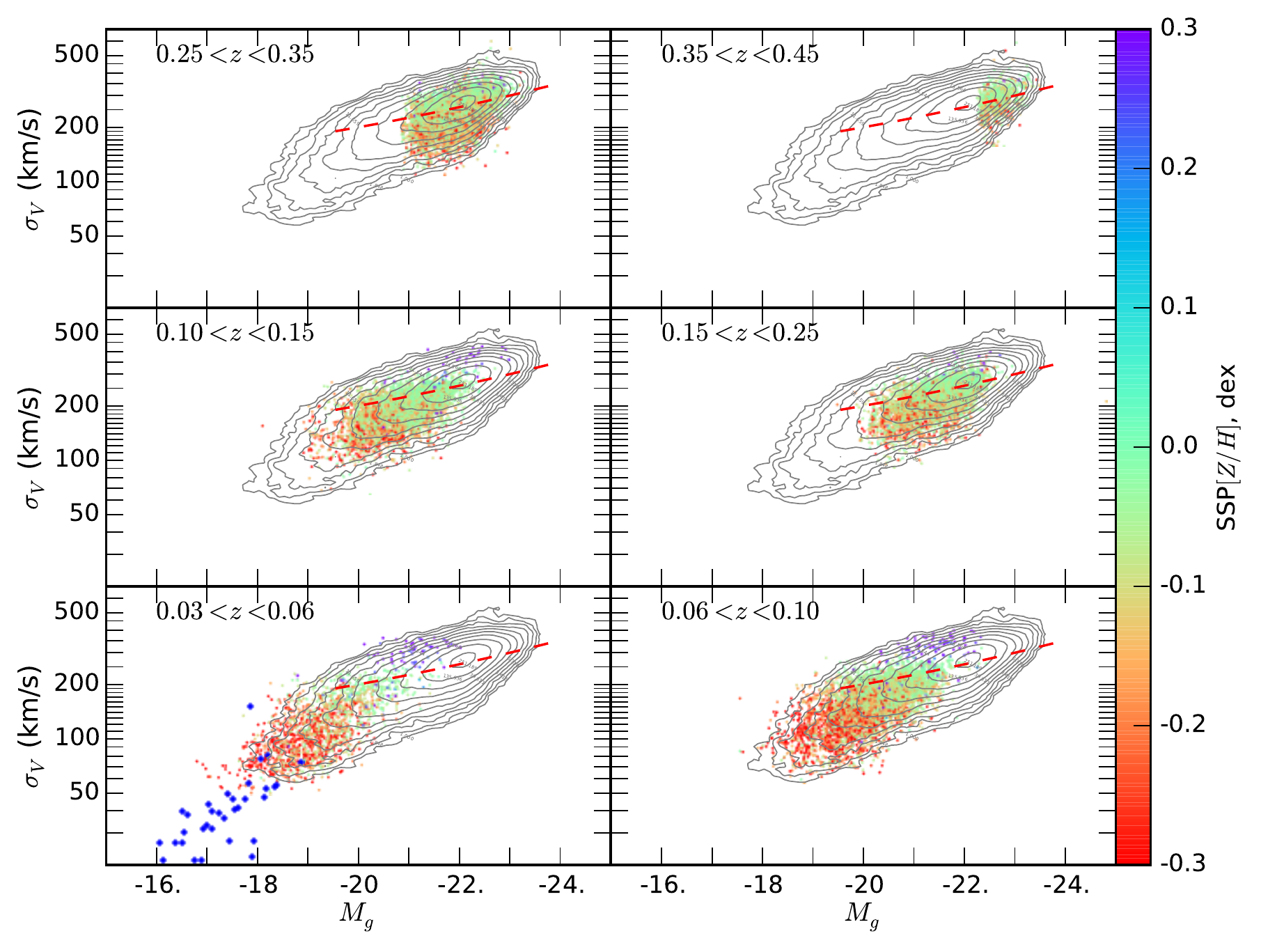}
\caption{Faber--Jackson relation for 52,506 morphologically classified elliptical 
galaxies in our sample (Galaxy Zoo classification).  In order to remove outliers, a 10$\%$ cut has been applied on the relative errors on g magnitude and on the velocity dispersion and a good adjustment has been required $\chi^2<0.8$. The contours correspond to the whole sample (smoothed with a 3$\times$3-pixels window). Each panel displays a redshift range corresponding to Figure \ref{fig_red_sequence}. The color coding corresponds to the SSP metallicity $[Z/H]$ displayed in Figure \ref{fig_ssp_vs_expSFH}. The dashed red line corresponds to the maximum likelihood estimate of the slope $L_g\propto \sigma^{4.00}$ at $z=0$ computed by \citet{2003AJ....125.1849B}. The blue points in the low redshift sub-sample (in the bottom left panel) correspond to the dwarf galaxies sample of  \citet{Chilingarian+08}. \label{figFJR}}
\end{figure*}

In Fig.~\ref{figFJR} we present the relation between galaxy luminosities and
velocity dispersions or the Faber--Jackson (\citeyear{FJ76}) relation
constructed for 52,506 elliptical galaxies which were morphologically
selected by the Galaxy Zoo \citep{Lintott+11} citizen science project and
had statistical uncertainties of their velocity dispersion measurements
better than 10\% of the value. We have corrected velocity dispersion
measurements to their global values according to \citet{Cappellari+06} using
half-light radii from \citet{Simard+11} included in our catalog.  We used
the criterion formulated in \citet{SMZC13}: in order to be included in our
early type galaxy sample, an object has to be classified by at least 10
Galaxy Zoo users of whom at least 70\% classify it as an elliptical
galaxy. Six panels present measurements in six different redshift intervals
shown as dots while the contours display the cumulative distribution at all
redshifts. The lowest redshift panel contains the measurements for a sample
of galaxies in the Abell~496 cluster ($z=0.033$) obtained from the analysis
of intermediate resolution (R=6300) spectra collected with the
FLAMES--Giraffe spectrograph at the 8-m Very Large Telescope of the European
Southern Observatory \citep{Chilingarian+08}. This dataset comprises mostly
dwarf early-type galaxies and it clearly forms an extension of the
low--luminosity part of the relation formed by the SDSS galaxies which
demonstrates that our velocity dispersion measurements at the low end do not
suffer from systematic errors connected to the spectral line spread
function uncertainties. The red dashed line represents the Faber--Jackson
relation for giant elliptical galaxies $L_g\propto \sigma^{4.00}$ at $z=0$ 
presented in \citet{2003AJ....125.1849B}. We see that the slope changes to 
$L_g\propto \sigma^{2.00}$ at fainter luminosities $M_g > -19.5$~mag similar
to what was demonstrated for a sample of dwarf galaxies in the Coma galaxy 
cluster by \citet{MG05}.

Because of the correlation of a galaxy luminosity with a stellar velocity
dispersion and the magnitude limited input galaxy sample, higher redshift
galaxies contribute only to the bins at high stellar velocity dispersions. 
Our catalog probes dwarf galaxies ($\sigma<100$~km~s$^{-1}$) at low
redshifts ($0.007<z<0.06$) that includes hundreds of massive galaxy clusters
and groups.

RCSED velocity dispersion measurements were used prior to
publication by \citet{SMZC13} for the calibration of the Fundamental Plane
\citep{DD87}.  We refer to that work for an intensive discussion regarding
the FP of elliptical galaxies observed by the SDSS.

\subsection{Stellar populations from absorption line analysis}

\begin{figure}
\includegraphics[width=\hsize]{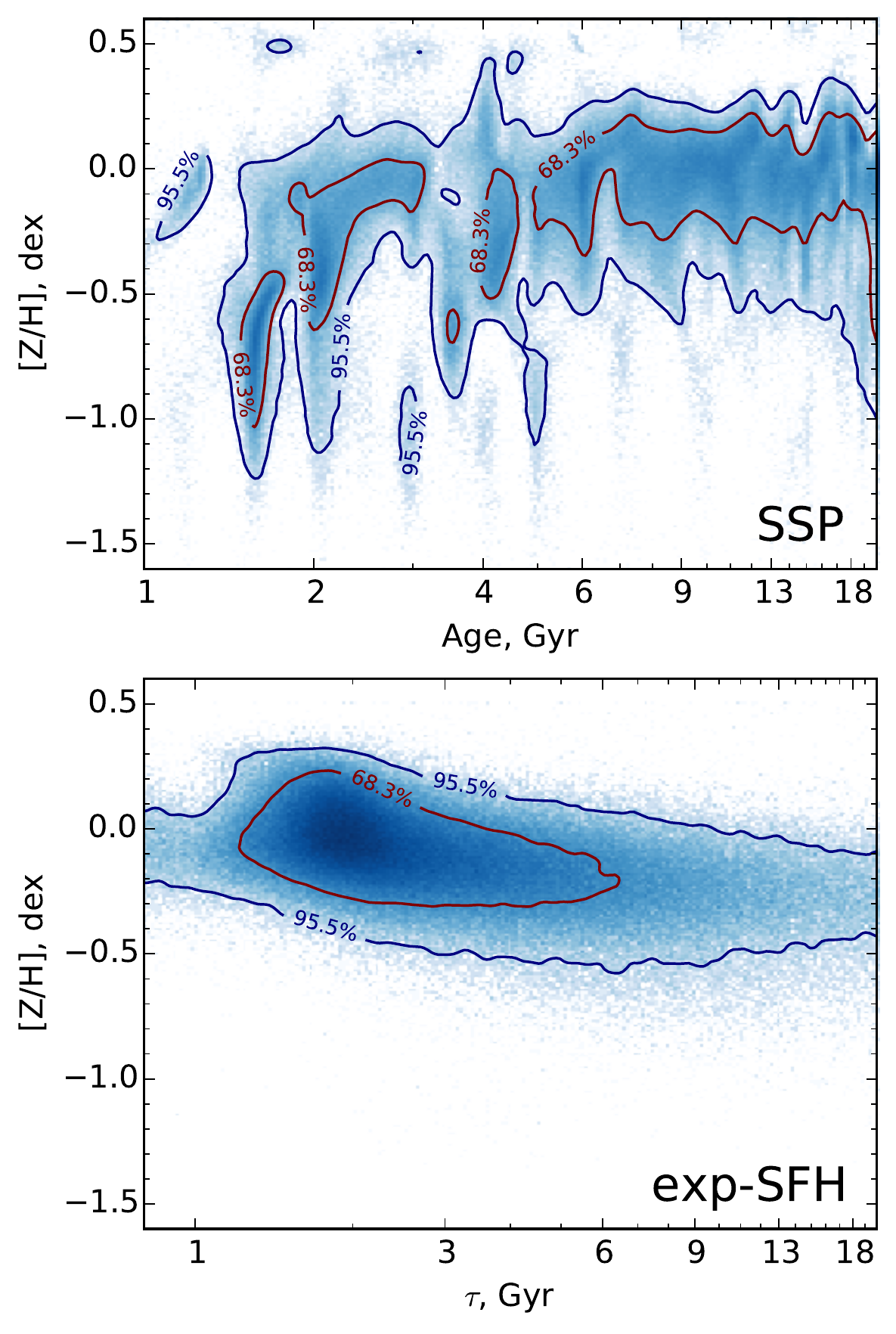}
\caption{Distributions of galaxies in the age--metallicity space from the
fitting of SSP (top panel) and exponentially decaying SFH (bottom panel)
models. \label{fig_age_met_SSP}}
\end{figure}

In our catalog we include stellar population parameters obtained by
the fitting of galaxy spectra using two stellar population model grids computed with the {\sc
pegase.hr} evolutionary synthesis code: (i) SSP models based on the
intermediate resolution MILES stellar library characterized by ages ($t$)
and metallicities ([Fe/H]); and (ii) models with exponentially declining
SFHs based on the high resolution ELODIE-3.1 stellar library characterized
by exponential timescales ($\tau$) and metallicities ([Fe/H]).  In
Fig~\ref{fig_age_met_SSP} we present distributions of galaxies in the two
parameter spaces.

One can clearly see a spotty structure in the SSP best fitting results and
the lack of such a structure for the exponentially decaying models.  We also
performed similar tests for original MILES stellar population models by
\citet{Vazdekis+10} and models by \citet{BC03a} for a sub-sample of SDSS DR7
spectra. We will provide a complete description and detailed discussion in
the forthcoming paper (Katkov et al. in prep.), here we present a brief
summary and conclusions of our study.

The observed spotty structure represents artifacts caused by the improper
implementation of the interpolation algorithm in the stellar population code
most likely on the stellar library interpolation step propagating into
stellar population models and not by the {\sc nbursts} population fitting
procedure.  The {\sc nbursts} code uses a non-linear minimization technique
that requires the second partial derivatives on all parameters to be
continuous.  Discontinuities will cause the solution to be either attracted
to some region of the parameter space, or pushed away from it.

Our conclusion is supported by the following observations: (i) the
morphology of the spotty structure remains very similar when using two
different sets of SSP models computed with the same {\sc pegase.hr} code but
with different stellar libraries, MILES and ELODIE; (ii) switching to
original MILES models \citep{Vazdekis+10} where the interpolation procedure
is much simpler than in {\sc pegase.hr} (linear interpolation between 5
nearest neighbors) changes the structure completely and strengthens the
artifacts; (iii) using exponentially decaying star formation models which
are constructed of numerous weighted SSPs removes most of the pattern at
$\tau \gtrsim 1.0$~Gyr but the structure still holds at $\tau < 1.0$~Gyr
where the number of co-added SSPs is small; (iv) smoothing a grid of {\sc
pegase.hr} MILES based SSP models using basic splines ($b$-splines) on age 
removes most of the pattern.

We also notice, that extending the working wavelength range to shorter
wavelengths ($<4500$~\AA) and increasing the multiplicative polynomial
degree strengthens the pattern while leaving spot positions in the pattern virtually
the same.  Therefore, we chose a very low order 5-th degree polynomial
continuum and restricted the wavelength range to $\lambda > 4500$~\AA\ for
the SSP fitting that produced stellar population parameters presented in our
catalog.

\begin{figure}
\includegraphics[width=\hsize]{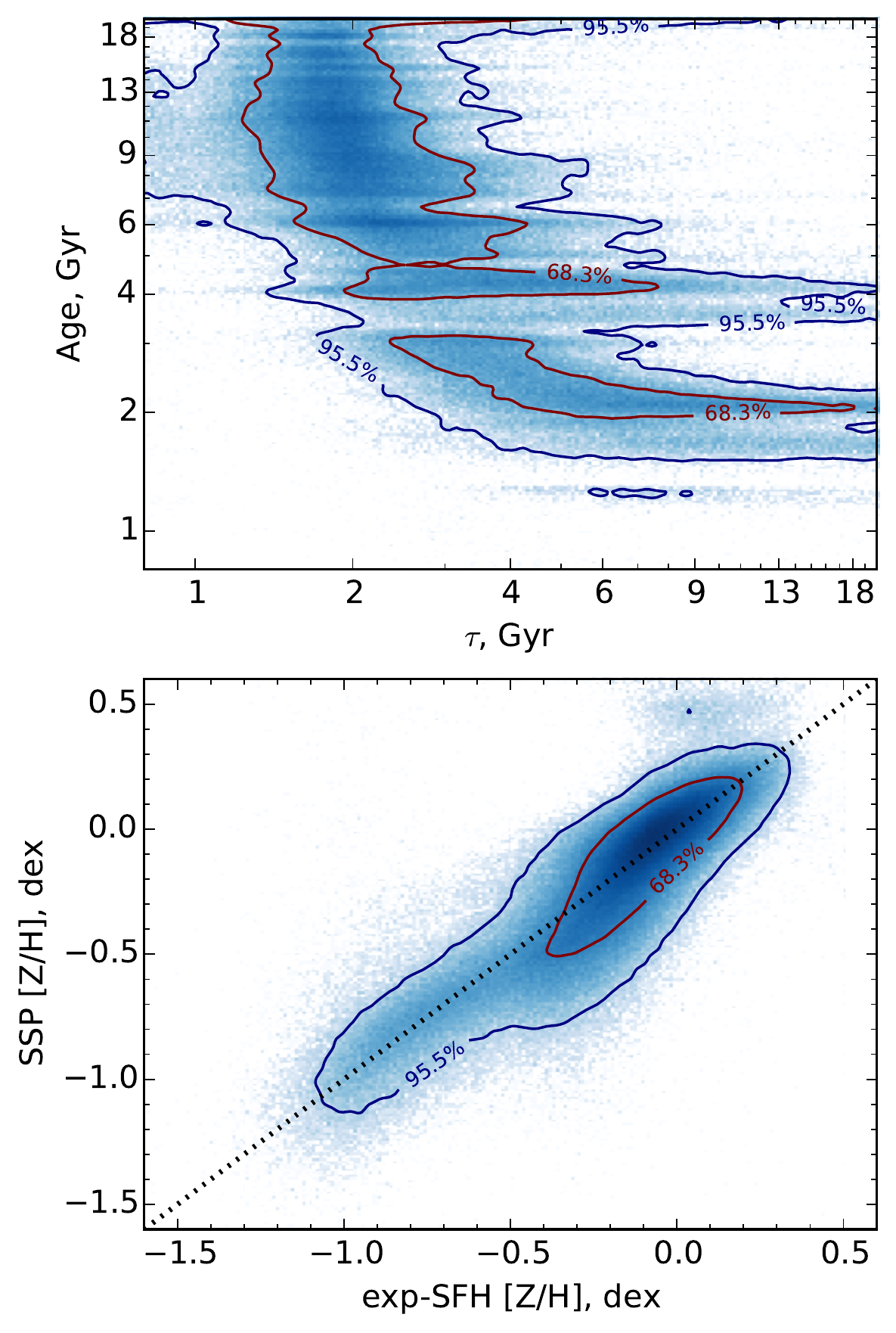}
\caption{Comparison of SSP ages to timescales $\tau$ for exponentially
decaying models (top) and metallicity measurements (bottom).
\label{fig_ssp_vs_expSFH}}
\end{figure}

In the top panel of Fig.~\ref{fig_ssp_vs_expSFH}, we present the comparison
of SSP ages and exponential timescales $\tau$. Despite the artifact
structure in ages that extends into horizontal stripes on this plot, there is
a 1-to-1 correspondence between $t$ and $\tau$ in a wide range of ages.
Short timescales $\tau$ correspond to old stellar populations while $\tau =
20$~Gyr is equivalent to $t \approx 1.8$~Gyr. The relation ``saturates'' for
younger stellar populations because they cannot be represented by
exponentially declining SFHs starting at high redshifts: either a later start
or multiple star formation episodes are needed to describe them.
\citet{CZ12} demonstrated that exponentially declining SFHs much better
represent observed broadband optical and UV colors than SSP models. In our
current sample about 16\% of galaxies ($\sim$131,500) have stellar populations too
young too be described by exponentially declining SFHs.

The bottom panel of Fig.~\ref{fig_ssp_vs_expSFH} displays the comparison of
stellar metallicities for the two sets of models. The agreement is very good
with a slight systematic difference between $-0.6<$[Fe/H]$<-0.2$~dex which we
attribute to the degeneracy between the metallicity and velocity dispersion
measurements for intermediate signal-to-noise ratios.

\begin{figure}
\includegraphics[width=\hsize]{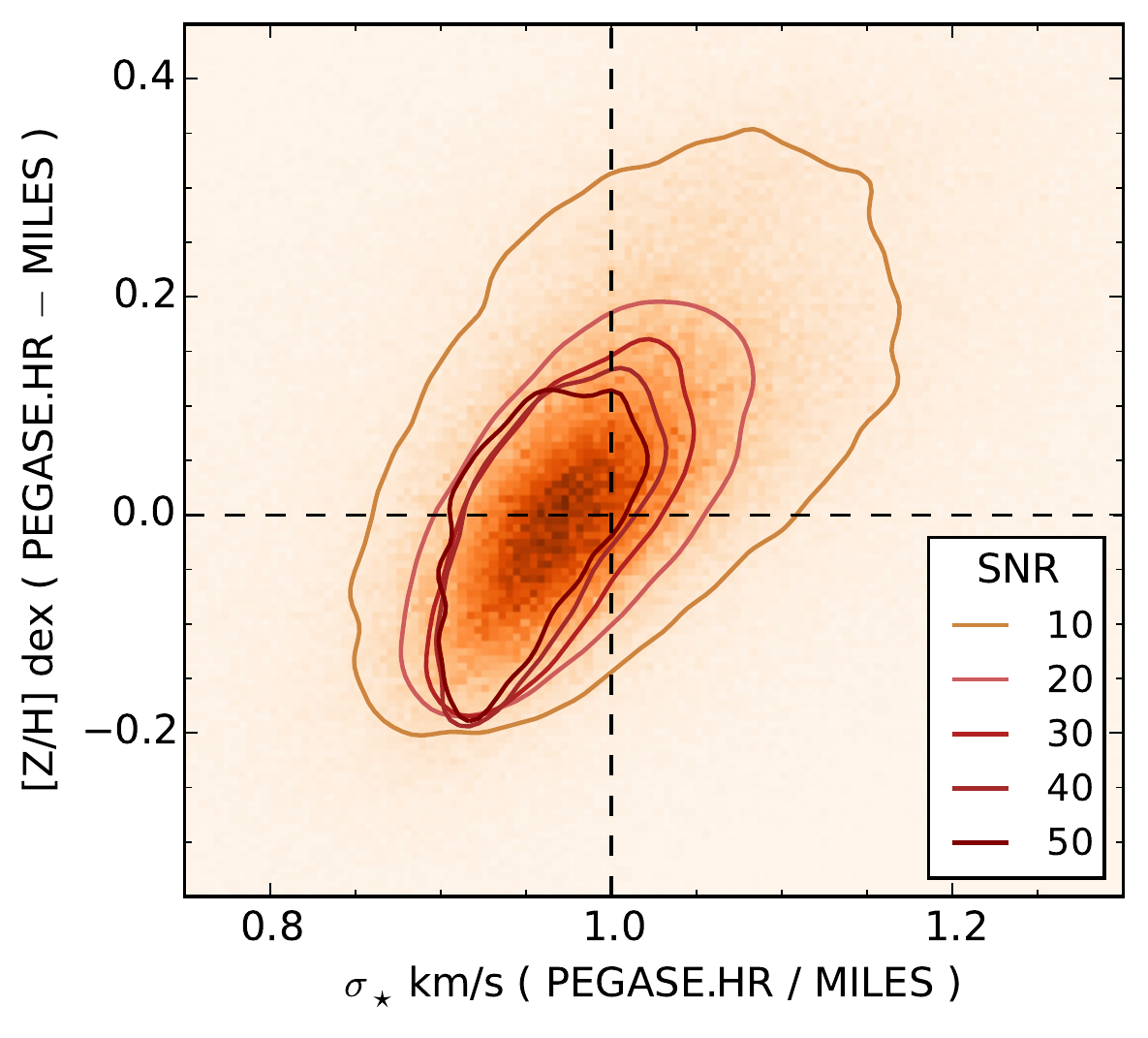}
\caption{Degeneracy between metallicity and velocity dispersion estimates
shown as the ratio between velocity dispersions vs SSP metallicities
obtained from the fitting of some 420,000 SDSS spectra using {\sc pegase.hr}
and MILES--PEGASE SSP models. The contours display the 1$\sigma$ values
which correspond to the areas containing 68\% galaxies with the
spectra having signal-to-noise ratios within 20\% of a displayed
value. The number of galaxies for each contour ranges from $\sim$1800
(S/N=50) to $\sim$111,000 (S/N=10).
\label{fig_degen}}
\end{figure}

We clearly see a substantial degeneracy between the metallicity and velocity
dispersion estimates which was pointed out in \citet{CPSA07}. In order to
perform a clear test, free of any effects connected to the usage of different
star formation histories, we fitted a subset of $\sim$420,000 spectra using
{\sc pegase.hr} SSP models in the wavelength range 3910--6790~\AA\ and
compared the metallicity and velocity dispersion measurements to those
obtained from the fitting of MILES--{\sc pegase} models against the same spectra.
In Fig.~\ref{fig_degen} we present the relation between the differences of
velocity dispersions and SSP metallicities obtained using the two sets of
models at different signal-to-noise ratios. We can clearly see the
degeneracy manifested by the elongated shape of the cloud that decreases
with the increasing signal-to-noise ratio up-to the signal-to-noise of 30.
Above 30 the improvement becomes insignificant. This result suggest that
published velocity dispersion values obtained with the full spectral fitting
of intermediate resolution spectra $R=1500-2500$ are subject to serious
systematic errors reaching 15\% of the measured value.

\subsection{Emission line properties}

\subsubsection{Comparison of line fluxes with the MPA--JHU and OSSY catalogs
and between the two techniques}

\begin{figure*}
\includegraphics[width=\textwidth]{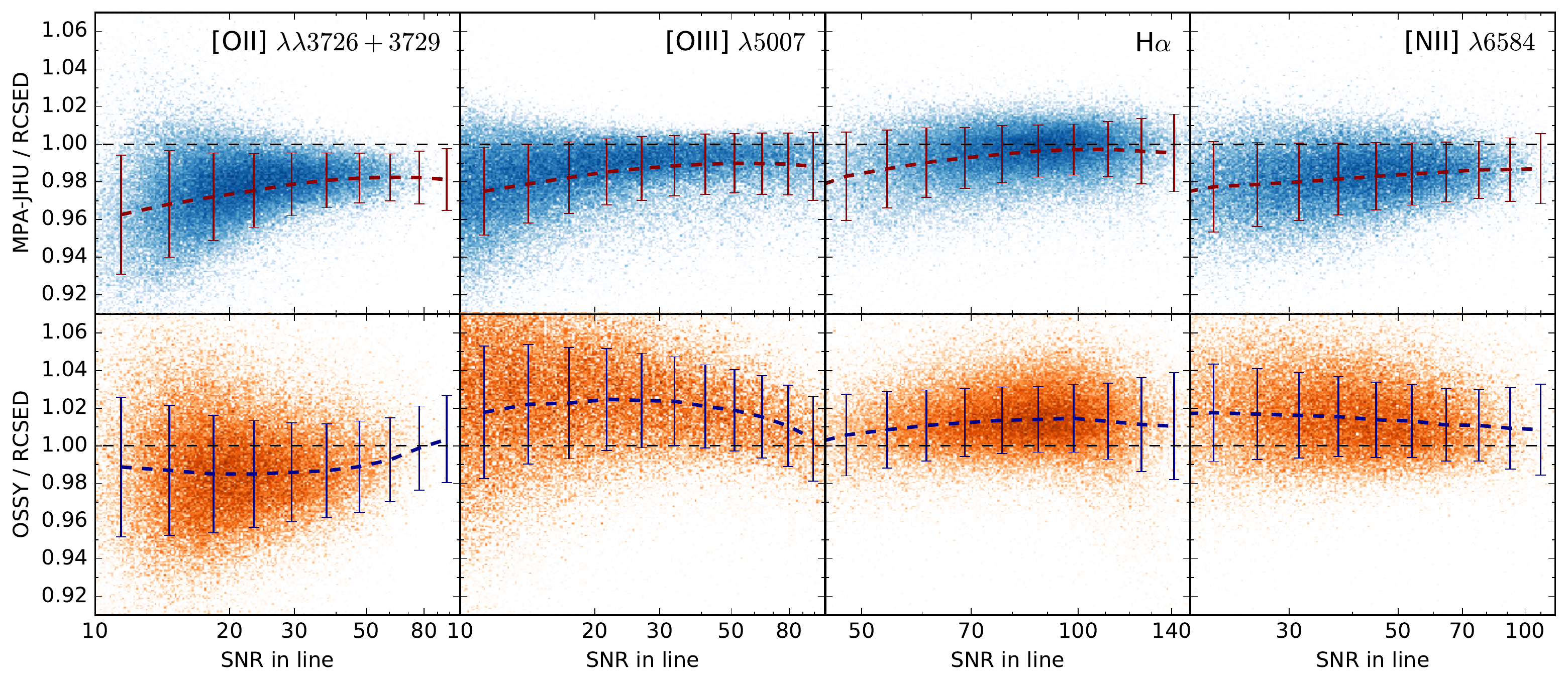}
\caption{Comparison of RCSED Gaussian emission line fluxes for 4 emission 
lines with the MPA--JHU (top row of plots) and OSSY (bottom row of plots) catalogs.
The median and standard deviation of the distribution are shown by brown
symbols with the error bars.}
\label{fig_emis_mpajhu_comp}
\end{figure*}

We compare a subset of our catalog containing measurements of emission
line fluxes obtained from the parametric Gaussian fitting to the results
from the MPA--JHU catalog distributed by the SDSS project
\citep{Brinchmann+04,Tremonti+04} and with the OSSY catalog
\citep{OSSY11}. We used OSSY emission line measurements prior to the internal
extinction correction. Given that the fluxes were computed using
very similar methodologies with the main different corresponding to the
subtraction of the underlying stellar population and the Galactic extinction
correction techniques used, we expect a very good agreement for well
detected emission lines. We directly compare fluxes of the
 [O{\sc ii}] (3727~\AA), [O{\sc iii}] (5007~\AA), H$\alpha$, and
[N{\sc ii}] (6584~\AA) emission lines for a sample of galaxies where they
were detected at a level exceeding 10$\sigma$ (i.e.  Flux/$\sigma($Flux$) >
10$).  The results are presented in Fig.~\ref{fig_emis_mpajhu_comp}.  We
obtain an excellent agreement with a systematic difference of less than
1\% and the standard deviation of residuals of about 2\% for
the bright end of the H$\alpha$ flux distribution.  At the faint end
(10$\sigma$ detection), the systematic difference stays within 2\%
while the standard deviation grows to 3\%.  Hence, we conclude that
our emission line fitting code works as expected and does not introduce any
substantial systematic errors to flux measurements.

Compared to H$\alpha$, the H$\beta$ line is much more sensitive to the
age of the stellar population being subtracted. In Appendix~\ref{sec_appsys} we discuss the
systematic errors of the H$\beta$ measurements as a function of the age mismatch.
In case of faint emission lines, the systematics dominates the measurements
if the age was determined incorrectly, and makes them useless for the emission 
line diagnostics.

\begin{figure}
\includegraphics[width=0.5\textwidth]{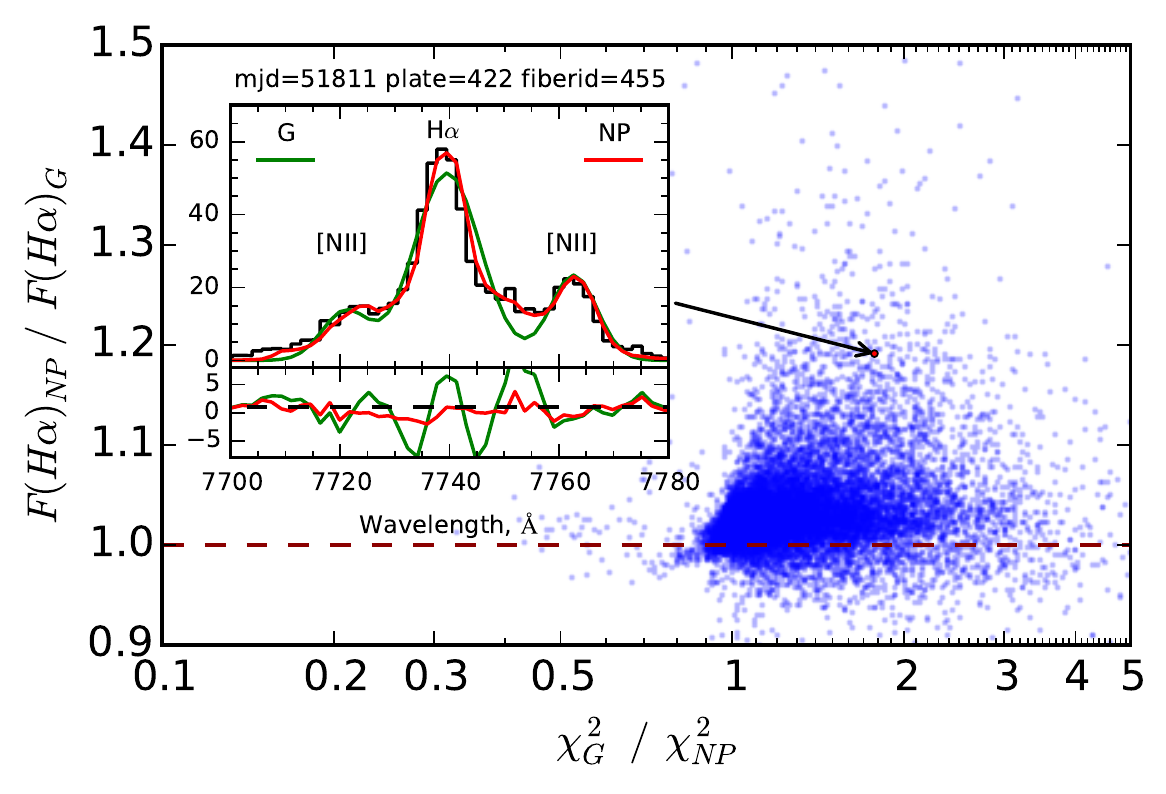}
\caption{Comparison of the H$\alpha$ fluxes obtained for the parametric (Gaussian) 
and non-parametric emission line profile fitting as a function of the $\chi^2$ ratio.
An example profile decomposition is shown in the inset for an object with highly 
discrepant flux estimates.}
\label{fig_emis_gaus_vs_np}
\end{figure}

The principal difference of our results with those published earlier is the
non-parametric approach to the emission line fitting.  For galaxies which
exhibit some signs of nuclear activity, the Balmer lines fluxes derived
non-parametrically significantly exceed the values obtained with the
Gaussian fitting.  In Fig.~\ref{fig_emis_gaus_vs_np} we present the
H$\alpha$ flux ratio between the two approaches.  The inset contains the
same Seyfert galaxy which we presented earlier in Fig.~\ref{fig_emis_lines}
and the arrow indicates its position in the diagram that suggests that its
non-parametric H$\alpha$ flux estimate is about 20\% higher than that
obtained with the Gaussian profile fitting.

\subsubsection{BPT diagrams and gas phase metallicities}

\begin{figure}
\includegraphics[width=\hsize]{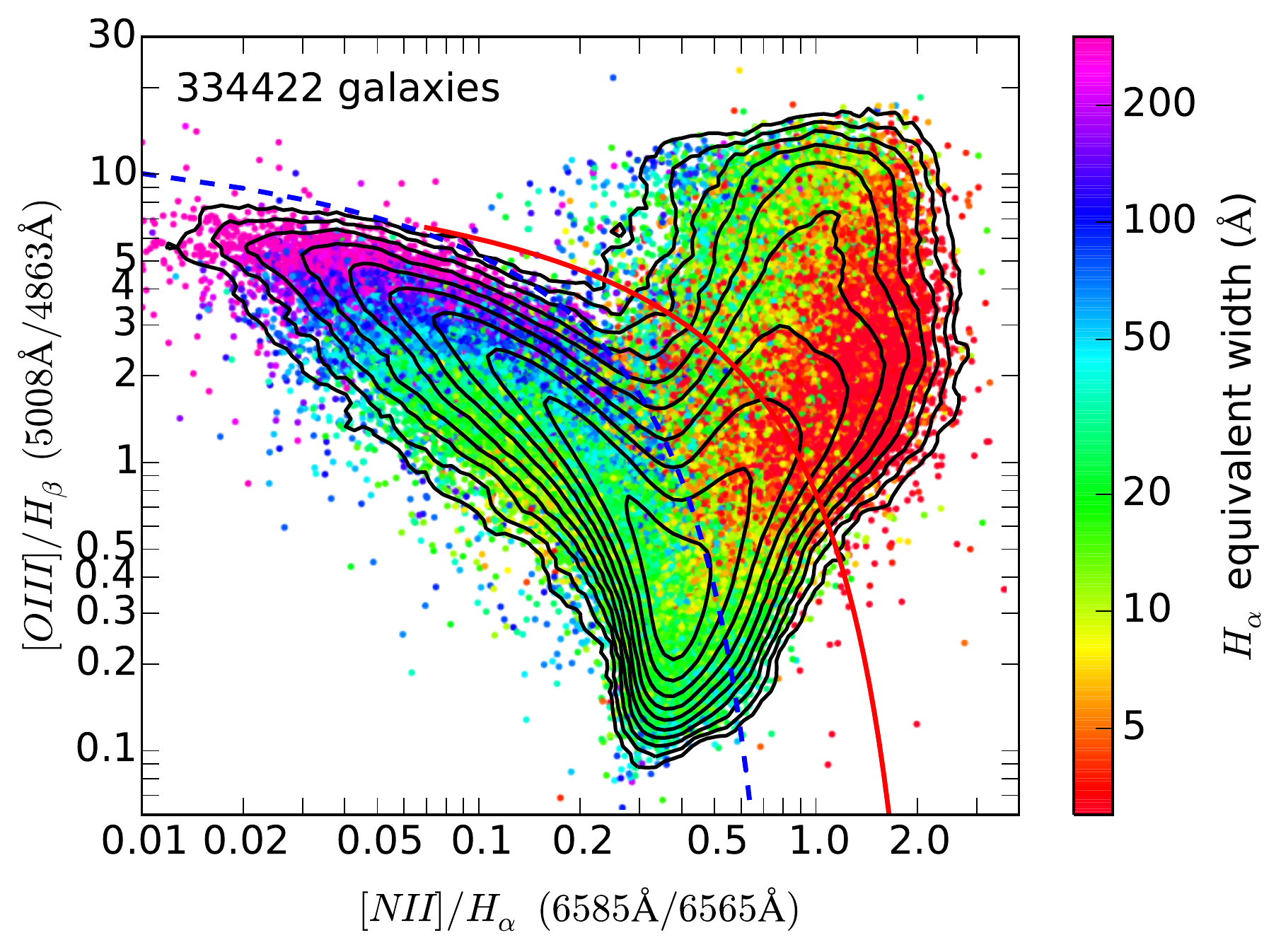}
\includegraphics[width=\hsize]{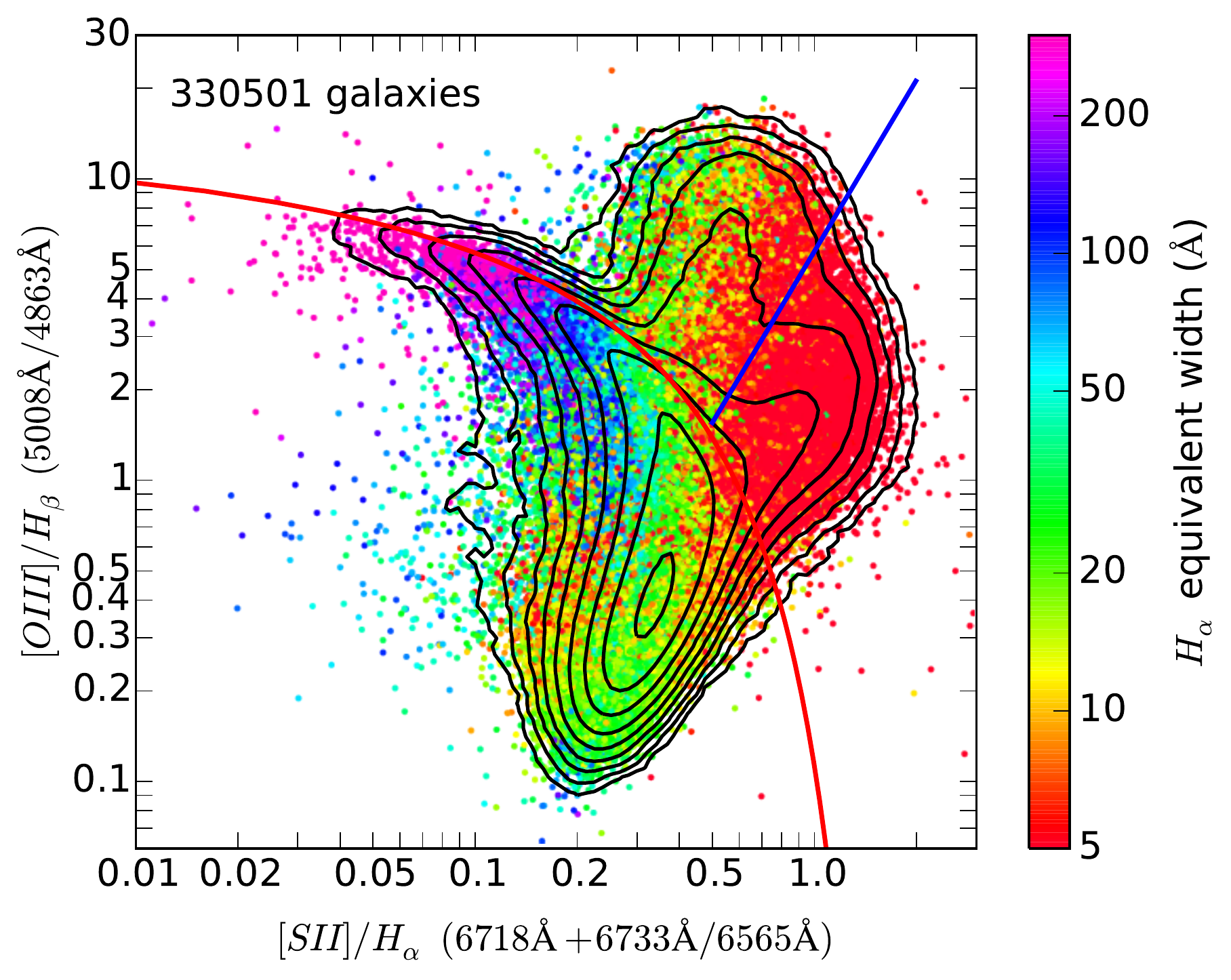}
\includegraphics[width=\hsize]{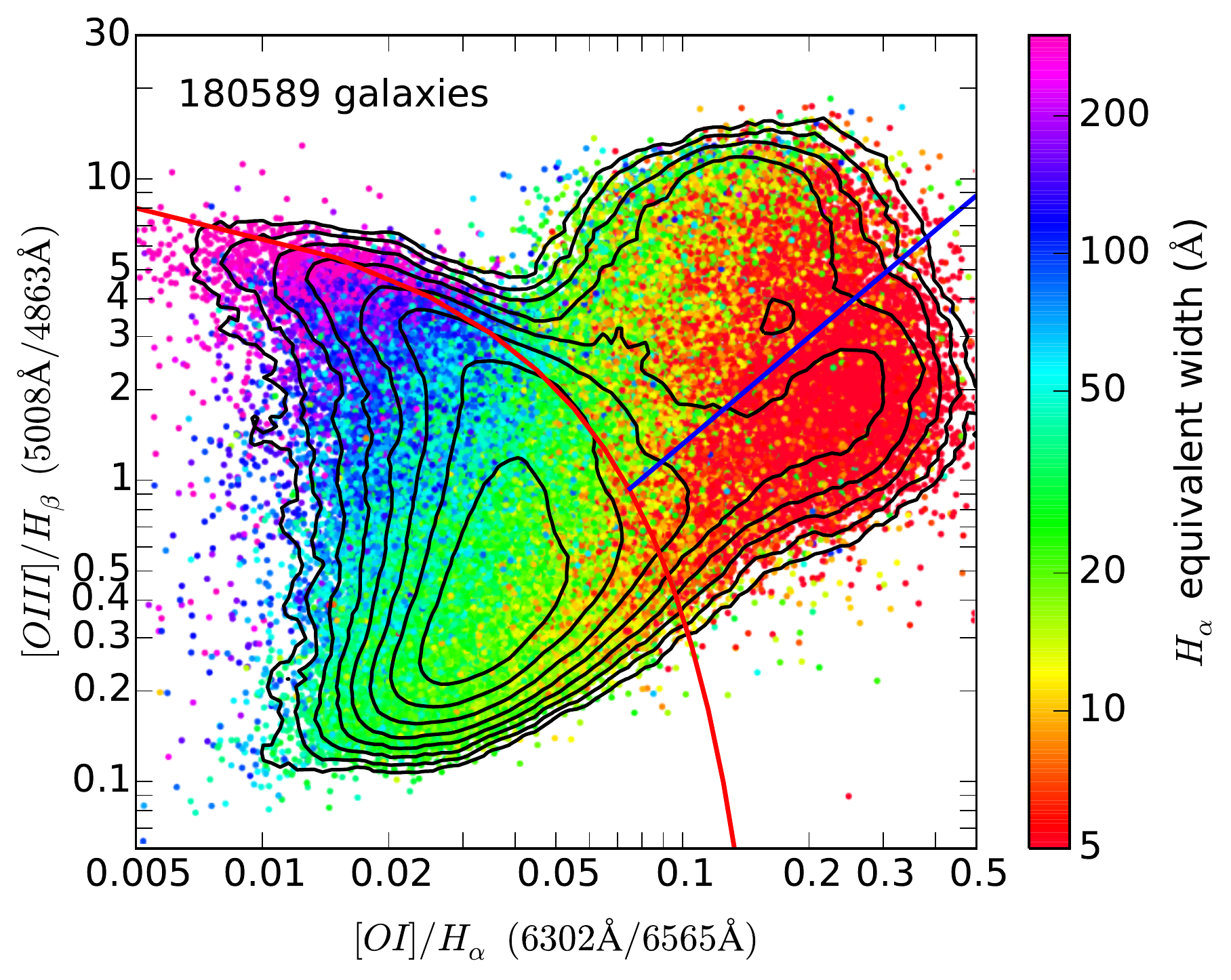}
\caption{Three flavors of a BPT diagram with the color coded $H_\alpha$ equivalent 
width. In each panel we display only those galaxies where 
all emission lines used in the corresponding plot have $S/N>3$. 
The contours correspond to the galaxy density smoothed with a moving average
based on a 4$\times$4-pixels window.  The number of galaxies kept in the
sample is indicated inside each panel.  The full and dashed lines correspond to
starforming and transitional galaxies in \citet{KGKH06}.}
\label{fig_BPTdiagram}
\end{figure}

In Fig.~\ref{fig_BPTdiagram} we present three flavors of the BPT diagram
which use different combinations of emission lines computed using the
non-parametric fitting.  The points are color-coded corresponding to the
H$\alpha$ emission line EW.  \citet{2010MNRAS.403.1036C} proposed to use the
H$\alpha$ EW to discriminate between Seyfert and LINER activity (instead of
the traditionally used [O{\sc iii}]/H$\beta$ ratio), because H$\beta$ is
often too weak to be detected and measured.  We clearly see the bimodal
distribution of non-starforming galaxies in the bottom two panels which
corresponds to Seyfert galaxies (cloud to the top) and
LINER/shockwave/post-AGB ionization (cloud to the bottom).  The top panel
displays the original BPT relation.  The region between the red solid and
the blue dashed lines defines ``transitional'' galaxies \citep{KGKH06} which
we included in the calculation of metallicities in addition to the star
forming galaxies located to the bottom left of the blue dashed line.

As we described above, RCSED includes gas phase metallicity measurements
calculated with the Bayesian method implemented in the IZI software package
with the \citet{Dopita+13} model grid, which uses all available emission
lines in a spectrum; and a recent technique by \citet{DKSN16} that relies
on the [N/O] calibration and uses only 5 emission lines around H$\alpha$.

\citet{KE08} demonstrated that different emission line calibrations yield
largely inconsistent gas phase metallicity estimates when applied to the
same input dataset with the differences reaching 0.7~dex (5 times).  There
is currently no consensus in the astronomical community about which
calibrations produce more reliable metallicity estimates with arguments for
both direct \citep{AM13} and indirect \citep{LopezSanchez+12} methods.  Gas
and stellar metallicities also seem to strongly disagree \citep{YKG12}.  We
notice, however, that all emission line calibrations result in the gas phase
[O/H] mass--metallicity relations spanning much lower range of metallicities
than that of stellar metallicities for a given galaxy stellar mass range. 
All gas metallicity relations saturate at high metallicities and the
saturation occurs at different values \citep[see e.g.  fig.~10 in][]{AM13}. 
The highest range of metallicities is covered by the calibration used by
\citet{Tremonti+04} and provided in the MPA--JHU catalog.

\begin{figure}
\includegraphics[width=\hsize]{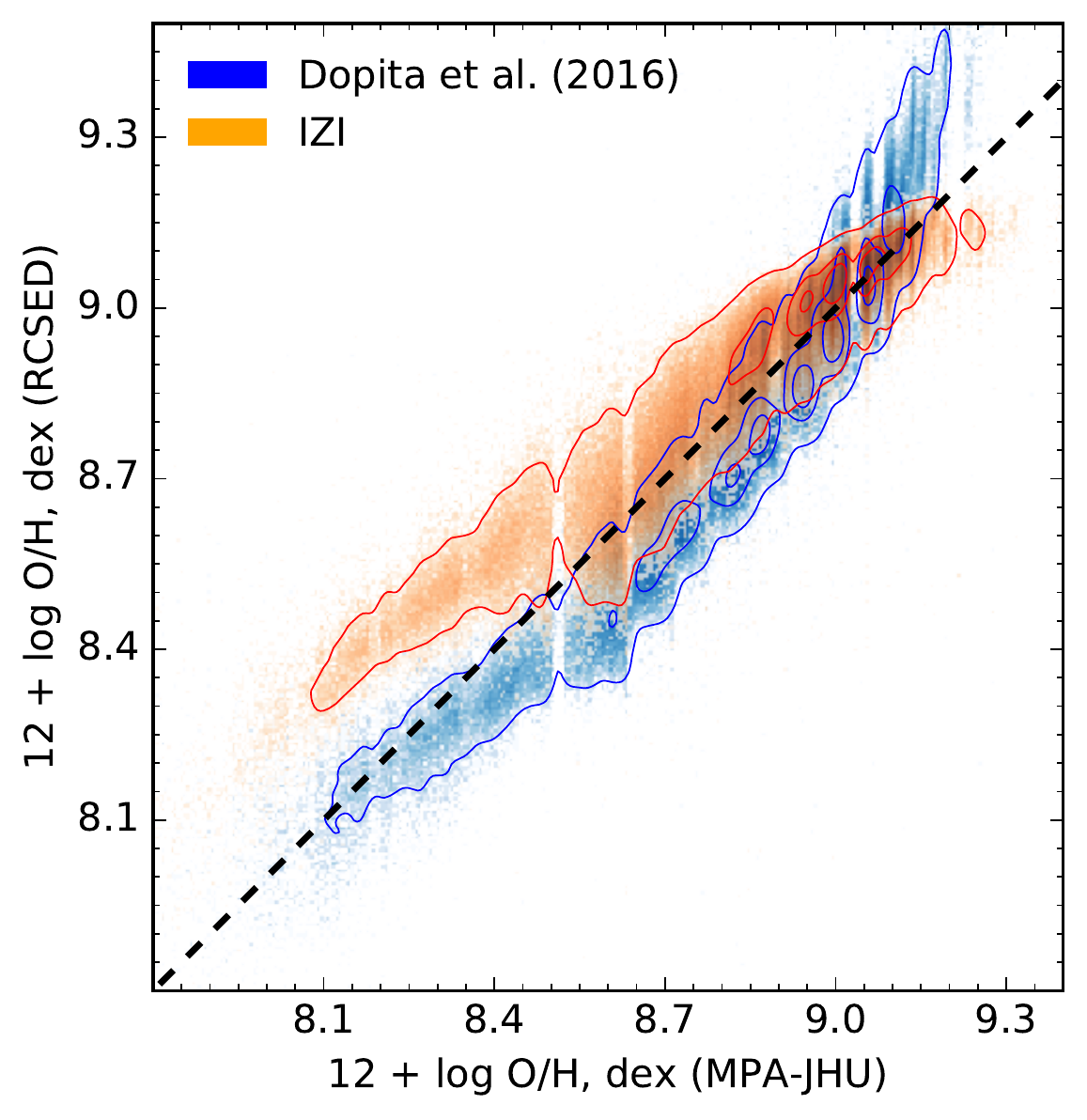}
\caption{Comparison of gas phase metallicities published in the MPA--JHU
catalog (horizontal axis) to our measurements (vertical axis). 
The results of the IZI Bayesian technique are shown in brown and the
measurements obtained with the \citet{DKSN16} calibration are shown in blue.
\label{fig_gas_met_comparison}}
\end{figure}

In Fig.~\ref{fig_gas_met_comparison} we present the comparison of the
\citet{DKSN16} calibration with the MPA--JHU metallicities (blue shaded
area) for 231,107 galaxies with the signal-to-noise ratios of H$\alpha$,
[N{\sc ii}], [O{\sc ii}], and [O{\sc iii}] lines exceeding 10.  The agreement is
very good at $12+$[O/H]$<9$~dex with the standard deviation of the
difference $0.08$~dex. At higher abundances \citet{DKSN16} metallicities become
slightly higher than those from the MPA--JHU dataset.

We ran the IZI metallicity determination code for a small sub-sample of 20,000
randomly selected starforming galaxies with high signal-to-noise emission
lines (S/N$>$10) using all available grids of models and compared the
derived metallicities with those obtained with the \citet{DKSN16}
calibration for the same galaxy sample.  The only model grid that
demonstrated a satisfactory agreement was that from \citet{Dopita+13}. As
expected, it also provides a satisfactory agreement with the MPA--JHU
catalog (see Fig.~\ref{fig_gas_met_comparison}, orange shaded areas) with
the standard deviation of the difference $0.10$~dex.

\begin{figure*}
\centerline{
\includegraphics[width=0.8\textwidth]{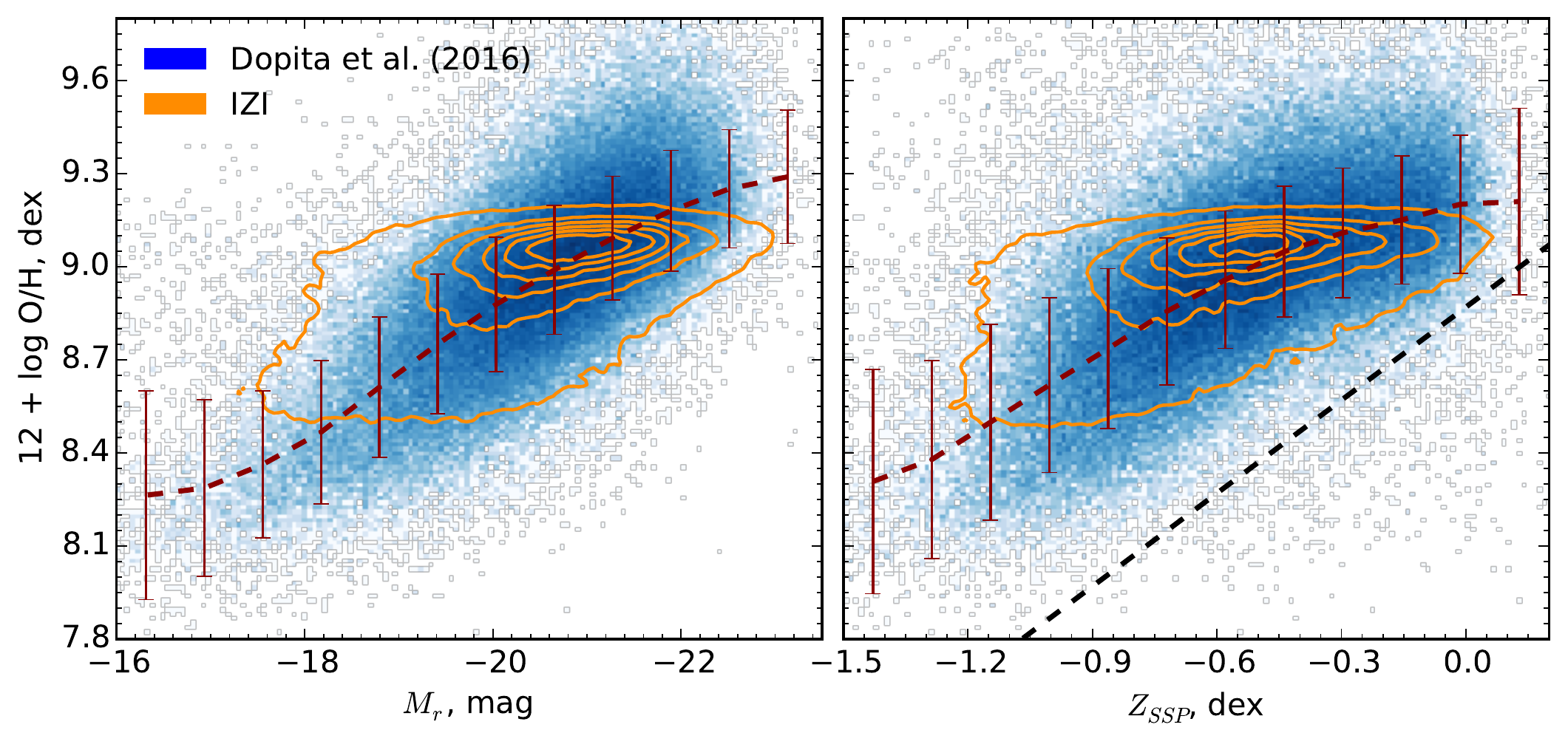}}

\caption{Relation between a gas phase metallicity and a galaxy luminosity (left) and an SSP stellar metallicity (right).
Blue dots and orange density contours display \citet{DKSN16} and IZI metallicities correspondingly.
The median and standard deviation of \citet{DKSN16} measurements are shown by brown symbols with the error bars.
\label{fig_emis_metallicities}}
\end{figure*}

In Fig.~\ref{fig_emis_metallicities} we show the luminosity--metallicity
relation (left panel) and the comparison of gas phase and SSP stellar
metallicities (right panel) for the IZI--based determination using the
\citet{Dopita+13} models (orange contours) and the \citet{DKSN16}
calibration (blue points).  The mass--metallicity relation is well defined
and we clearly see that the \citet{Dopita+13} model grid used in IZI yields
a flatter shape than the more recent calibration \citep{DKSN16}.

The comparison of gas phase and stellar metallicities reveal a substantial offset
ranging from about 0.3~dex at solar stellar metallicities to 0.8~dex at the
low end ([Fe/H]$_{\mathrm{star}}=-1.1$~dex).  Keeping in mind that stellar and gas phase
metallicities might have different zero points and should not be directly
compared to each other, the observed pattern is exactly what is expected due
to the self enrichment of stellar populations happening in galaxies with
extended star formation histories.  The stars during their evolution form
heavy elements which then get ejected into the ISM and recycled in the
subsequent generations of stars, hence, increasing their metal abundances
\citep[see e.g.][]{Matteucci94}.  Therefore, younger generations of stars
become more metal rich.  SSP models probe mean
stellar metallicities over the entire lifetime of a galaxy weighted with the
stellar $M/L$ ratios and the star formation rate while the gas phase
metallicity reflects the current chemical abundance pattern in the ISM
enriched with metals, therefore, we expect to see the offset in
metallicities. On the other hand, for the constant metal production rate per
solar mass, the difference at low metallicities will be higher because the
metallicity scale is logarithmic, therefore stellar metallicities should
span a larger range of value compared to gas phase metallicities and the
mass--metallicity relation slopes for gas will be shallower than that for
stars.

\section{Catalog access: web-site and Virtual Observatory access
interfaces}

Efficient, convenient, and  intuitive data access mechanisms and interfaces
are essential for a complex project like RCSED. Therefore, we
decided to build access interfaces for both interactive and batch access to
the data.

\begin{figure}
\includegraphics[width=\hsize]{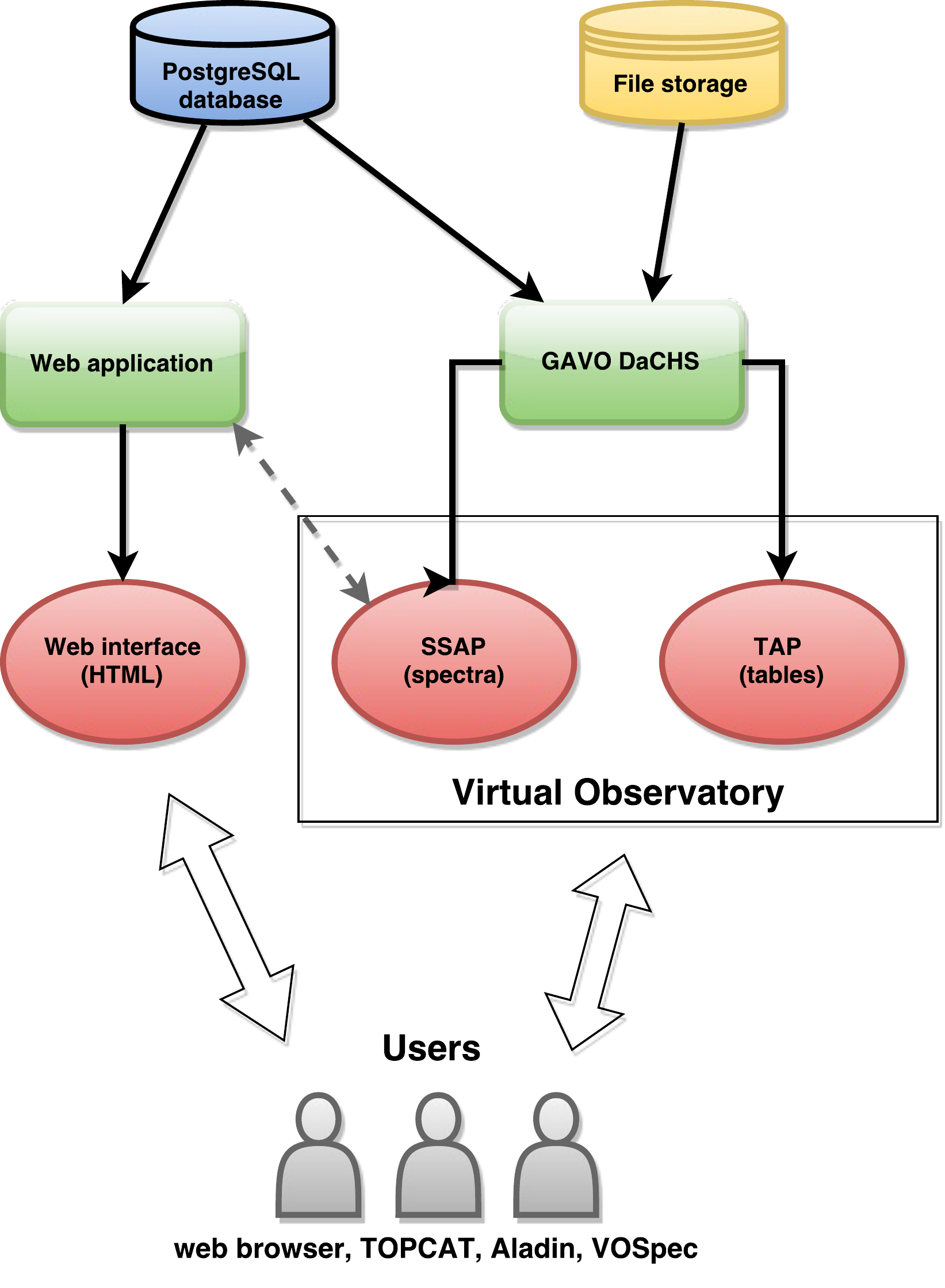}
\caption{Block diagram of the catalog data access infrastructure.
Data are stored in the relational database (catalog tables and spectra metadata) and on the disk (FITS files with spectra and continuum models).
They are accessed by applications (a custom web application and the GAVO DaCHS suite) which in turn expose several public access interfaces suitable for convenient queries and data retrieval by the multitude of user client programs, both VO-compatible and generic.
\label{fig_block_diagram}}
\end{figure}

RCSED includes several different data types (e.g. spectra and tabular data)
and our access infrastructure (see Fig.~\ref{fig_block_diagram}) is
organized to simplify their usage through different interfaces.  The most
natural way to access the catalog is by using the web application at
\url{http://rcsed.sai.msu.ru/}.  It provides a single-field {\sc
google}-style search interface where one can query the catalog by an object
identifier, coordinates or object properties, e.g. \emph{select all galaxies with
redshifts $z<0.1$ having red colors $g-r>1.5$}.  Every object in the
sample has its own web page with the summary of all its properties, SED,
spectral data available in the catalog, and image cutouts displaying the
object at different wavelength provided by GALEX, SDSS, and UKIDSS surveys. 
An example of a spectrum summary plot presented on such web pages for every
object is given in Fig.~\ref{fig_emspec}. 

We developed an Application 
Programming Interface (API) to UKIDSS data, which allow us to extract image
cutouts around an arbitrary position with a given box size in every filter. 
From cutout images in the \emph{JHK} bands we generate a color composite
image and display it in the object web-page. The API implemented in {\sc
python} is available for download from the project web-site. Another service 
we present is an interactive spectrum plotter implemented in JavaScript,
our alternative to the SDSS spectrum plotter.  It contains a number of
value-added features, such as the display of best-fitting templates and
identification of emission lines.

In addition to the custom web application, our data distribution infrastructure
has the open source GAVO DaCHS\footnote{\url{http://soft.g-vo.org/dachs}} data center
suite in its core (see Fig.~\ref{fig_block_diagram}) which provides a set of
VO data access mechanisms.

The data for SDSS spectra and their best-fitting SSP models are provided as
FITS files that can be fetched by direct unique URLs.  One can find a
URL for every particular object spectrum file either in the object's web page or by
querying the provided IVOA Simple Spectral Access Protocol (SSAP) web service using object
coordinates. The SSAP web service answers essentially with a list of spectra
URLs and it is convenient to access programmatically or by using 
VO compatible client applications such as
TOPCAT\footnote{\url{http://www.star.bris.ac.uk/~mbt/topcat/}}
\citep{Taylor05},
SPLAT-VO\footnote{\url{http://www.g-vo.org/pmwiki/About/SPLAT}} or
VO-Spec\footnote{\url{http://www.sciops.esa.int/index.php?project=SAT&page=vospec}}
which can directly load spectral data for further analysis by analyzing
the SSAP web service query result.

\begin{figure*}
\includegraphics[width=\hsize]{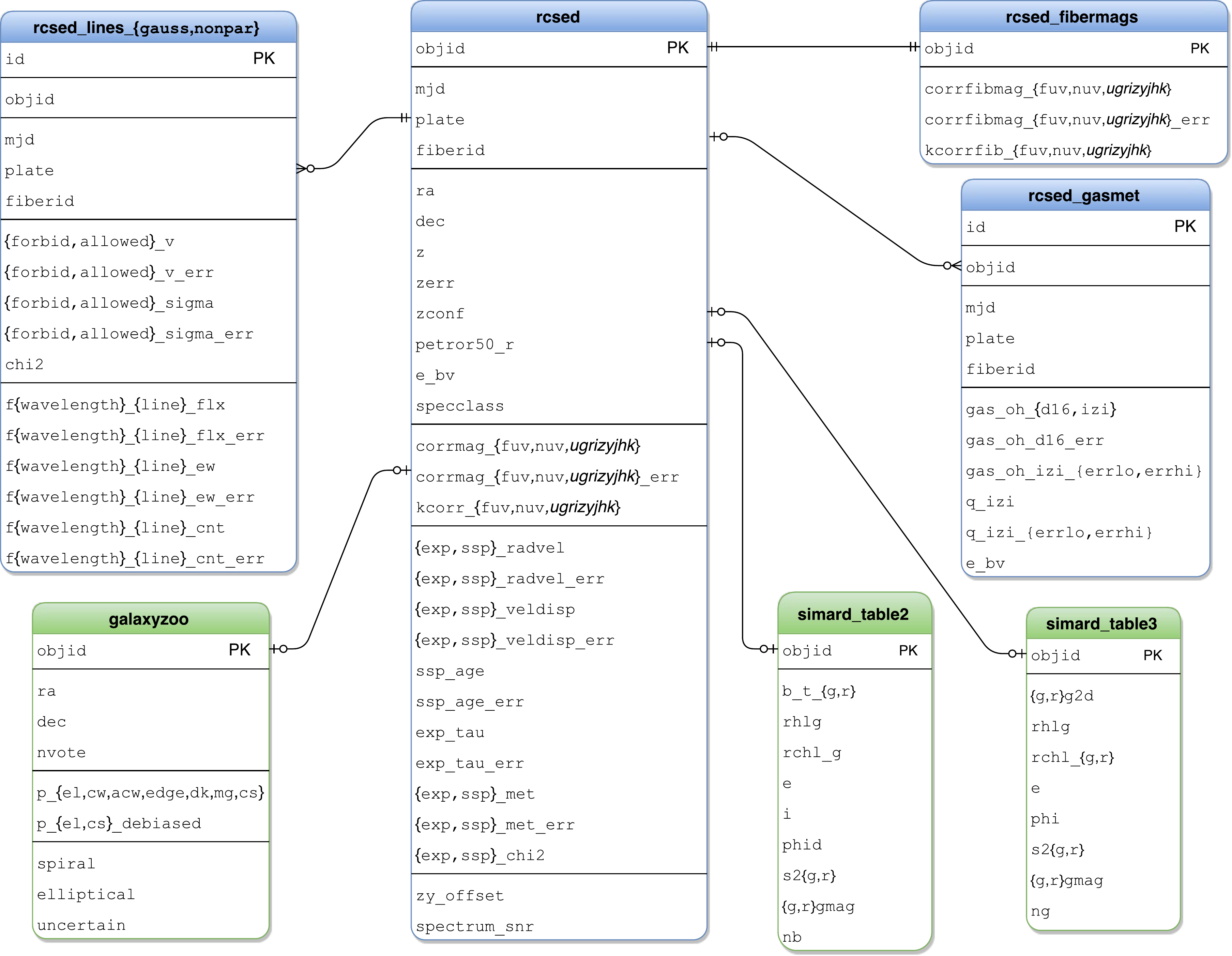}
\caption{The entity--relationship diagram for the tables in the catalog database.
Blue color denotes original tables computed in RCSED, green color is for external datasets added to the database for convenience.
Main table is {\tt rcsed} which has a 1-to-1 relation to the {\tt rcsed\_fibermags} table by the primary key column {\tt objid},a 1-to-many (optional) relation to the {\tt rcsed\_gasmet} table with gas phase metallicity measurements, 1-to-many (optional) relations to {\tt rcsed\_lines\_gauss}, {\tt rcsed\_lines\_nonpar} tables with parametric (gauss) and non-parametric emission lines measurements.
{\tt galaxyzoo}, {\tt simard\_table2} and {\tt simard\_table3} datasets are all linked to the main table by the {\tt objid} column and provide morphological classification and structural properties of galaxies from our sample using the data form  \citet{Lintott+11} and \citet{Simard+11} respectively.
All these tables are stored in {\tt specphot} database schema, so the properly qualified table name is, for example, {\tt specphot.rcsed}.
We also note that all column names in the database are lowercase for homogeneity.
Same schema and relationships applies to the distribution of RCSED in the form of FITS tables available for download from project's website.
\label{fig_er_diagram}}
\end{figure*}

For the ultimate flexibility of querying tabular catalog data, we provide a Table Access Protocol (TAP) web service.
IVOA TAP is an access interface, which allows a user to query the entire relational database schema (see Fig.~\ref{fig_er_diagram}) using a powerful SQL-like language.
It can be considered as an open source equivalent of the SDSS CasJobs service.
Again, TAP web service can be used for script-based access as well as by using desktop VO applications.
In particular, TOPCAT has a very useful TAP query dialogue with built-in help, query examples, syntax highlighting and given database schema assistance tools.
We encourage our users to access the RCSED TAP web service through TOPCAT.
We also note that our TAP service has a table upload capability, so that the user can upload his/her own tables and use it in subsequent SQL queries i.e. in {\sc join} clauses, that is convenient for cross-identification of user provided object samples with the RCSED objects without the need of downloading our full catalog. 

Below we give several query examples that are helpful to start using the RCSED database.
More query examples and science case tutorials are provided on the project website \url{http://rcsed.sai.msu.ru}.
Our TAP web service can be used for joining tables from the database schema {\tt specphot}
presented in Fig.~\ref{fig_er_diagram}, so that it is easy to retrieve a
single table with the GalaxyZoo morphology, the photometric bulge+disk 
decomposition and the RCSED basic parameters combined for any galaxy of interest.
An example of such a query to select all those data for a particular object would be:

\begin{verbatim}
SELECT 
  r.*, g.*, s2.*
FROM 
  specphot.rcsed AS r 
  JOIN specphot.galaxyzoo AS g 
    ON r.objid = g.objid
  JOIN specphot.simard_table2 AS s2
    ON r.objid = s2.objid
WHERE
  r.objid = 587731891649052703
\end{verbatim}

Note that {\tt specphot} prefix for table names corresponds to the name of the database schema where RCSED tables are stored.
A query to retreive all data from the RCSED on galaxies, classified as ellipticals in GalaxyZoo is:

\begin{verbatim}
SELECT
  r.*
FROM
  specphot.rcsed AS r
  JOIN specphot.galaxyzoo AS g 
    ON r.objid = g.objid
WHERE
  g.elliptical = 1
\end{verbatim}

A query to select the data for a BPT \citep{BPT81} diagram for 10,000 galaxies with ${\rm S/N} > 10$ in the corresponding line fluxes obtained with the Gaussian fitting looks like this:

\begin{verbatim}
SELECT 
  TOP 10000 
  f6550_nii_flx / f6565_h_alpha_flx AS BPT_x, 
  f5008_oiii_flx / f4863_h_beta_flx AS BPT_y 
FROM 
  specphot.rcsed_lines_gauss 
WHERE 
  f6565_h_alpha_flx / f6565_h_alpha_flx_err > 10 
  AND f5008_oiii_flx / f5008_oiii_flx_err > 10 
  AND f4863_h_beta_flx / f4863_h_beta_flx_err > 10 
  AND f6550_nii_flx / f6550_nii_flx_err > 10
\end{verbatim}

Finally, all the catalog tables (see Fig.~\ref{fig_er_diagram}) are available for download as FITS tables from the project's website for the offline use.

\section{Summary}

We presented a reference catalog of homogeneous multi-wavelength
spectrophotometric information for some 800,000 low to intermediate redshift
galaxies ($0.007<z<0.6$) from the SDSS DR7 spectroscopic galaxy sample with
value-added data. For every galaxy we provide: 
\begin{itemize} 
\item a $k$-corrected and Galactic extinction corrected far-UV to NIR broad
band SED for integrated fluxes compiled from the SDSS (optical), GALEX (UV),
and UKIDSS (NIR) surveys
\item a  $k$-corrected and Galactic extinction corrected far-UV to NIR broad
band SED for fluxes in circular 3-arcsec apertures that correspond to SDSS
spectral apertures
\item results of the full spectrum fitting of an SDSS spectrum using the
{\sc nbursts} technique that includes: (a) an original SDSS spectrum; (b)
the best-fitting simple stellar population template in the wavelength
range $3700<\lambda<6800$~\AA\ and the best-fitting stellar population model with
an exponentially declining star formation history in the wavelength range
$3900<\lambda<6800$~\AA; (c) estimates of stellar radial velocities,
velocity dispersions, age, an exponential characteristic timescale for the
star formation history, metallicities for two sets of stellar population models
\item results of the emission line analysis using parametric (Gaussian) and
non-parametric line profiles that include: (a) emission line fluxes
corrected for the Galactic extinction; (b) estimates of the reddening inside
a galaxy for star formation dominated systems derived from the observed Balmer
decrement; (c) radial velocity offsets with respect to stars; (d) intrinsic 
emission line widths for the parametric fitting
\item cross-match of a galaxy with third-party catalogs providing its
structural parameters from the two-dimensional light profile fitting
\citep{Simard+11} and galaxy morphology by the Galaxy Zoo project
\citep{Lintott+08}
\end{itemize}

The catalog is fully integrated into the international Virtual Observatory
infrastructure and available via a web application and as a Virtual
Observatory resource providing IVOA TAP and IVOA SSAP interfaces in order to
programmatically access tabular data and spectra correspondingly.

In addition to that, we presented best-fitting polynomial approximations for
the red sequence shape in color--magnitude diagrams that include different
colors, and mean colors for galaxies of 6 morphological types, from
elliptical to late type spirals and irregulars and 3 luminosity classes
(giants, intermediate luminosity, dwarfs).

Our catalog has already been used in several research projects that can be
categorized into two groups: (i) statistical studies of galaxy properties;
(ii) search and discovery of rare galaxies. 

The first interesting result obtained with RCSED was the discovery of a
universal 3-dimensional relation of \emph{NUV} and optical galaxy colors and
luminosities \citep{CZ12}.  It also demonstrated that the integrated
\emph{NUV}$-$\emph{r} color is a good proxy for a morphological type. The
spectum fitting results for elliptical galaxies were later used in the
re-calibration of the fundamental plane in SDSS \citep{SMZC13} which allowed
us to compute redshift independent distances to early-type galaxies.
Finally, we performed a calibration of near-infrared stellar $M/L$ ratios
using optical colors and computed stellar masses for a new catalog of groups
and clusters combining SDSS and 2MASS redshift survey data
\citep{Saulder+16}. Our non-parametric emission line fitting results will be
used to perform massive determinations of virial black hole masses in AGNs
(Katkov et al. in prep). Other potential applications of statistical studies
based on RCSED include but not limited to: environmental dependence of
galaxy scaling relations and stellar population properties; connecting
AGNs to stellar populations in galaxy centers; comparing different star
formation rate indicators (e.g. emission line fluxes, UV and MIR
photometry).

Thanks to the unique combination of photometric and spectral data as
well as physical properties of galaxies derived from them, RCSED becomes an
efficient search tool for rare or unique galaxies.  Our dataset was used to
discover and characterize massive compact galaxies at intermediate redshifts
$0.2<z<0.8$ \citep{DCHG13} which were thought to exist only in the early
Universe ($z>1.5$) and measure their volume density \citep{DHGC14}.  Then,
using their fundamental plane positions, intermediate redshift compact
galaxies were shown to be an extension of normal ellipticals to the compact
regime \citep{ZDGC15}.  Finally, it was demonstrated that some massive
compact early-type galaxies actually stopped forming stars very recently
\citep{Zahid+16}.  We also used the universal UV--optical
color--color--magnitude relation to define complex selection criteria and
discover 195 previously considered extremely rare compact elliptical
galaxies \citep{CZ15}.  One can identify other obvious extragalactic
rarities easily searchable with RCSED: post-starburst galaxies, candidate
double-peaked AGNs, dwarf AGN hosts, ``normal'' galaxies with peculiarities
detectable in multi-wavelength data such as ellipticals with NUV excess.

In the future, we anticipate to release intermediate and high redshift
extensions of our catalog that will include the analysis of publicly
available spectra from the Smithsonian Astrophysical Observatory Hectospec
archive\footnote{\url{http://oirsa.cfa.harvard.edu/}} collected with the
Hectospec multi-fiber spectrograph \citep{Fabricant+05} the 6.5-m MMT and
the DEEP2 galaxy redshift survey \citep{Newman+13} made with the DEIMOS
spectrograph at the 10-m Keck telescope.  We also plan to expand the
wavelength coverage by adding the all-sky infrared data from the
Wide-field Infrared Survey Explorer (WISE) satellite \citep{Wright+10}.
A major update to our catalog will be made with the full
spectrophotometric fitting the entire sample using the {\sc nbursts+phot}
algorithm \citep{CK12} and resolving star formation histories for about
$10^5$ galaxies with high quality UV and NIR data.

\section*{Acknowledgments} 

We acknowledge our anonymous referee whose comments helped us to improve
this manuscript.
IC's reseach is supported by the Smithsonian Astrophysical Observatory
Telescope Data Center.  IZ acknowledges the support by the Russian
Scientific Foundation grant 14-50-00043 for the catalog assembly
tasks and grant 14-12-00146 for the data publication and deployment system.
The authors acknowledge partial support from the M.V.Lomonosov
Moscow State University Program of Development, and a Russian--French PICS
International Laboratory program (no.  6590) co-funded by the RFBR (project
15-52-15050), entitled ``Galaxy evolution mechanisms in the Local Universe
and at intermediate redshifts''.  The statistical studies of galaxy
populations by IC, IZ, IK, and ER are supported by the RFBR grant
15-32-21062 and the presidential grant MD-7355.2015.2.  The authors are
grateful to citizen scientists M.~Chernyshov, A.~Kilchik, A.~Sergeev,
R.~Tihanovich, and A.~Timirgazin for their valuable help
with the development of the project website.  In 2009--2011 the project was
supported by the VO-Paris Data Centre and by the Action Specifique de
l'Observatoire Virtuel (VO-France).  A substantial progress in our project
was achieved during our 2013, 2014, and 2015 annual Chamonix workshops and
we are grateful to our host O.~Bevan at Ch\^alet des Sapins.
This research has made use of TOPCAT, developed by Mark Taylor at the
University of Bristol; Aladin developed by the Centre de Donn\'ees
Astronomiques de Strasbourg (CDS); the ``exploresdss'' script by G.~Mamon
(IAP); the VizieR catalogue access tool (CDS).  Funding for the \emph{SDSS}
and \emph{SDSS}-II has been provided by the Alfred P.  Sloan Foundation, the
Participating Institutions, the National Science Foundation, the U.S. 
Department of Energy, the National Aeronautics and Space Administration, the
Japanese Monbukagakusho, the Max Planck Society, and the Higher Education
Funding Council for England.  The \emph{SDSS} Web Site is
\url{http://www.sdss.org/}.  \emph{GALEX} (Galaxy Evolution Explorer) is a
NASA Small Explorer, launched in April 2003.  We gratefully acknowledge
NASA's support for construction, operation, and science analysis for the
\emph{GALEX} mission, developed in cooperation with the Centre National
d'Etudes Spatiales of France and the Korean Ministry of Science and
Technology.

\bibliographystyle{apj}
\bibliography{biblib}

\appendix

\section{Systematics in emission line measurements due to stellar population template mismatch}
\label{sec_appsys}

Absorption lines of the hydrogen Balmer series contain important
information about stellar population ages \citep{Worthey94}, they become
weaker when stars get older.  At the same time, emission Balmer
lines are used for the ISM diagnostic and star formation studies
\citep{BPT81}.  For the vast majority of galaxies in our sample, we see
relatively weak emission lines on top of a stellar continuum.  Therefore, in
order to accurately measure emission line fluxes, we need to precisely model
stellar populations.  Hence, when gas emission lines reside on top of a
stellar continuum, any systematic uncertainty in the modelling of absorption
lines will affect emission line measurements. Specifically, the age mismatch in
the stellar population fitting will substantially bias Balmer line fluxes.

In order to quantify this effect, we performed the following procedure: (i)
We selected 2,000 spectra from our sample with Balmer emission line
intensities ranging from weak to strong based on their equivalent widths;
(ii) we fitted those spectra using stellar population model grids fixing the
SSP age to 2, 4, 8, and 16~Gyr; (iii) we measured emission line fluxes in
the fitting residuals in these four sets of spectra; (iv) we compared them
to emission line fluxes obtained for best-fitting stellar populations
presented in our catalog.

\begin{figure}
\includegraphics[width=0.5\hsize]{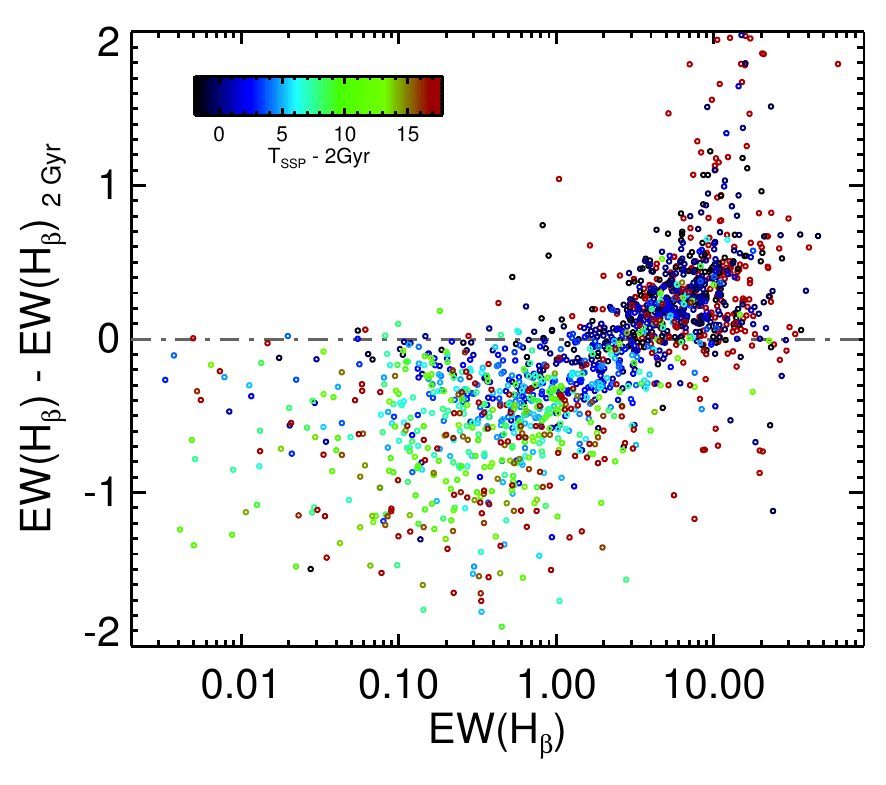}
\caption{The stellar population age mismatch effect on H$\beta$ flux
measurements. The difference of the H$\beta$ EW computed using the best
fitting SSP template and a template with the age fixed to 2~Gyr is plotted
against the measured H$\beta$ EW for the best fitting SSP template.
The age difference between the best fitting SSP age and 2~Gyr is color coded.
\label{fig_EMLAge}}
\end{figure}

In Fig.~\ref{fig_EMLAge} we present our results. It is clear, that the age
mismatch affects emission line fluxes for weak lines:
The systematic errors grow when lines become weaker, and the difference
between the best fitting and the fixed ages templates gets higher. When ages are
underestimated by the fitting procedure (i.e. a galaxy is older than the
age of a template), Balmer emission line fluxes are underestimated too.
Because forbidden lines often used in the gas state diagnostics (e.g. [N{\sc
ii}] or [O{\sc iii}]) do not lie on top of strong age sensitive absorption
features, their fluxes remain virtually unaffected, hence, moving a galaxy
over the diagnostic plots (e.g. BPT) and potentially leading to the
ionization mechanism misclassification.

\section{Catalog compilation: SQL query}
\label{sec_sql}

When selecting the core sample of galaxies we performed the following SQL query in the SDSS CasJobs service in the DR7 context (see details in Section~\ref{sec_sample}):

\begin{verbatim}
SELECT
     p.objID, p.ra, p.dec, 

     p.modelMag_u, p.modelMagErr_u, p.modelMag_g, p.modelMagErr_g, 
     p.modelMag_r, p.modelMagErr_r, p.modelMag_i, p.modelMagErr_i,
     p.modelMag_z, p.modelMagErr_z, 

     petroMag_u, petroMagErr_u, petroMag_g, petroMagErr_g, 
     petroMag_r, petroMagErr_r, petroMag_i, petroMagErr_i,
     petroMag_z, petroMagErr_z,

     p.fiberMag_u, p.fiberMagErr_u, p.fiberMag_g, p.fiberMagErr_g, 
     p.fiberMag_r, p.fiberMagErr_r, p.fiberMag_i, p.fiberMagErr_i, 
     p.fiberMag_z, p.fiberMagErr_z,

     p.petroR50_u, p.petroR50Err_u, p.petroR50_g, p.petroR50Err_g, 
     p.petroR50_r, p.petroR50Err_r, p.petroR50_i, p.petroR50Err_i,
     p.petroR50_z, p.petroR50Err_z, 

     p.extinction_u, p.extinction_g, p.extinction_r, p.extinction_i, p.extinction_z,

     s.specObjID, s.mjd, s.plate, s.fiberID, 
     s.z, s.zerr, s.zconf, s.objType, s.sn_0, s.sn_1, s.sn_2, 
     (SELECT stripe FROM dbo.fCoordsFromEq(p.ra,p.dec)) AS stripe, 
     s.specClass

INTO mydb.RCSED_SDSS
FROM PhotoObj AS p, SpecObj as s
WHERE
    s.bestObjid = p.objID
    AND s.z >= 0.007
    AND s.z < 0.6
    AND s.specClass IN (dbo.fSpecClass('GAL_EM'), dbo.fSpecClass('GALAXY'))
\end{verbatim}

This query returned 800,311 rows with 12 duplicate objects for which SDSS {\tt SpecObj} table contains 2 records despite it is documented to be clean from duplicates.
We discard these duplicate spectra by keeping the record with higher S/N out of each pair of duplicates (and hence having e.g. better redshift estimate).
From now we continue with the sample of 800,299 galaxies.

The coordinates of obtained galaxies were then uploaded to the GALEX CasJobs service and the following query was performed there in GALEXGR6Plus7 context:

\begin{verbatim}
SELECT
     sdss.objid, 
     galex_objid,
     
     nuv_mag, nuv_magerr, fuv_mag, fuv_magerr,
     nuv_mag_aper_1, nuv_magerr_aper_1, nuv_mag_auto,
     fuv_mag_aper_1, fuv_magerr_aper_1, fuv_mag_auto,
     e_bv
INTO mydb.RCSED_SDSS_GALEX

FROM
     (
          SELECT
               s.objid, 
               (SELECT objid FROM dbo.fGetNearestObjEq(s.ra, s.dec, 0.05)) AS galex_objid
          FROM
               mydb.RCSED_SDSS_coords AS s
     ) AS sdss
JOIN
     photoObjAll AS p
     ON sdss.galex_objid = p.objid
\end{verbatim}

This query returned 485,996 rows.

\label{lastpage}

\section{Catalog column descriptions}

In Tables~\ref{tab_metadata_main}--\ref{tab_metadata_nonpar} we provide descriptions and metadata for columns of the original tables of RCSED, which are shown in blue in Fig.~\ref{fig_er_diagram}.
The external datasets available in the RCSED database are described in the corresponding original papers (see the text for references).

This column information is identical for FITS tables distribution of the catalog, as well as when accessing the RCSED database through the Table Access Protocol, or using the catalog website \url{http://rcsed.sai.msu.ru}.
For each column name in every table we give: 
(i) units (dash sign indicates that a column is dimensionless or units are not applicable to it); 
(ii) data type in the database convention in order to guide a user on the precisionm and puropse of a column;
(iii) IVOA Unified Content Descriptor (UCD) that helps one to identify equivalent physical quantities available for comparison in the VO or to associate a column and its uncertainty; 
and (iv) human readable description of the column contents.
When a table includes many similar columns as in the case of spectral lines properties in the {\tt rcsed\_lines\_gauss} and {\tt rcsed\_lines\_nonpar} database tables, we only give metadata for first group of columns in it and abridge the rest (Table~\ref{tab_metadata_gauss} and Table~\ref{tab_metadata_nonpar}).
The complete list of emission lines included in our catalog  and the column name prefixes in {\tt rcsed\_lines\_gauss} and {\tt rcsed\_lines\_nonpar} are given in Table~\ref{tbl_linelist}.

\begin{table*}
\caption{Main catalog table ({\tt rcsed}) columns metadata and descriptions.
\label{tab_metadata_main}}
\begin{scriptsize}
\begin{tabular}{lcccp{7cm}}
Column & Units & Datatype & UCD & Description \\
\hline
 objid & - & bigint & meta.id;meta.main & SDSS ObjID (unique identifier) \\
 specobjid & - & bigint & meta.id & SDSS SpecObjID (unique identifier within spectral galaxies sample) \\
 mjd & - & integer & time.epoch & MJD of observation \\
 plate & - & smallint & meta.id & SDSS plate ID \\
 fiberid & - & smallint & meta.id & SDSS fiber ID \\
 ra & deg & double & pos.eq.ra;meta.main & RA (J2000) of galaxy \\
 dec & deg & double & pos.eq.dec;meta.main & Dec (J2000) of galaxy \\
 z & - & real & src.redshift & Galaxy redshift \\
 zerr & - & real & stat.error;src.redshift & Uncertainty of galaxy redshift \\
 zconf & - & real & stat.fit.param;src.redshift & SDSS $r$edshift confidence \\
 petror50\_r & arcsec & real & phys.angSize & SDSS $r$adius containing 50\% of Petrosian flux \\
 e\_bv & mag & real & phot.color.excess & E(B-V) at this (l,b) from SFD98 \\
 specclass & - & smallint & src.spType & SDSS spectral classification \\
 corrmag\_fuv & mag & real & phot.mag;em.UV.FUV & Galactic extinction corrected total (Kron-like elliptical aperture) magnitude in GALEX $FUV$ filter \\
corrmag\_nuv & mag & real & phot.mag;em.UV.NUV & Same as above for GALEX $NUV$ filter \\
corrmag\_u & mag & real & phot.mag;em.opt.U & Galactic extinction corrected total (Petrosian) magnitude in SDSS $u$ filter \\
corrmag\_g & mag & real & phot.mag;em.opt.B & Same as above for SDSS $g$ filter \\
corrmag\_r & mag & real & phot.mag;em.opt.R & Same as above for SDSS $r$ filter \\
corrmag\_i & mag & real & phot.mag;em.opt.I & Same as above for SDSS $i$ filter \\
corrmag\_z & mag & real & phot.mag;em.opt.I & Same as above for SDSS $z$ filter \\
corrmag\_y & mag & real & phot.mag;em.IR.J & Same as above for UKIDSS $Y$ filter \\
corrmag\_j & mag & real & phot.mag;em.IR.J & Same as above for UKIDSS $J$ filter \\
corrmag\_h & mag & real & phot.mag;em.IR.H & Same as above for UKIDSS $H$ filter \\
corrmag\_k & mag & real & phot.mag;em.IR.K & Same as above for UKIDSS $K$ filter \\
corrmag\_fuv\_err & mag & real & stat.error;phot.mag;em.UV.FUV & Uncertainty of corrmag\_fuv column  \\
corrmag\_nuv\_err & mag & real & stat.error;phot.mag;em.UV.NUV & Uncertainty of  corrmag\_nuv column \\ 
corrmag\_u\_err & mag & real & stat.error;phot.mag;em.opt.U & Uncertainty of corrmag\_u column \\ 
corrmag\_g\_err & mag & real & stat.error;phot.mag;em.opt.B & Uncertainty of corrmag\_g column \\ 
corrmag\_r\_err & mag & real & stat.error;phot.mag;em.opt.R & Uncertainty of corrmag\_r column \\ corrmag\_i\_err & mag & real & stat.error;phot.mag;em.opt.I & Uncertainty of corrmag\_i column \\
corrmag\_z\_err & mag & real & stat.error;phot.mag;em.opt.I & Uncertainty of corrmag\_z column \\
corrmag\_y\_err & mag & real & stat.error;phot.mag;em.IR.J & Uncertainty of corrmag\_y column \\
corrmag\_j\_err & mag & real & stat.error;phot.mag;em.IR.J & Uncertainty of corrmag\_j column \\
corrmag\_h\_err & mag & real & stat.error;phot.mag;em.IR.H & Uncertainty of corrmag\_h column \\
corrmag\_k\_err & mag & real & stat.error;phot.mag;em.IR.K & Uncertainty of corrmag\_k column \\
kcorr\_fuv & mag & real & arith.factor;em.UV.FUV & K-correction for GALEX $FUV$ magnitude \\
kcorr\_nuv & mag & real & arith.factor;em.UV.NUV & Same as above for GALEX $NUV$ magnitude \\
kcorr\_u & mag & real & arith.factor;em.opt.U & K-correction for (Petrosian) SDSS $u$ magnitude \\
 kcorr\_g & mag & real & arith.factor;em.opt.B & Same as above for SDSS $g$ magnitude \\
 kcorr\_r & mag & real & arith.factor;em.opt.R & Same as above for SDSS $r$ magnitude \\
 kcorr\_i & mag & real & arith.factor;em.opt.I & Same as above for SDSS $i$ magnitude \\
 kcorr\_z & mag & real & arith.factor;em.opt.I & Same as above for SDSS $z$ magnitude \\
 kcorr\_y & mag & real & arith.factor;em.IR.J & Same as above for UKIDSS $Y$ magnitude \\
 kcorr\_j & mag & real & arith.factor;em.IR.J & Same as above for UKIDSS $J$ magnitude \\
 kcorr\_h & mag & real & arith.factor;em.IR.H & Same as above for UKIDSS $H$ magnitude \\
 kcorr\_k & mag & real & arith.factor;em.IR.K & Same as above for UKIDSS $K$ magnitude \\
 exp\_radvel & km/s & real & spect.dopplerVeloc.opt & Radial velocity (exp SFH) \\
 exp\_radvel\_err & km/s & real & stat.error;spect.dopplerVeloc.opt & Radial velocity error (exp SFH) \\
 exp\_veldisp & km/s & real & phys.veloc.dispersion & Velocity dispersion (exp SFH) \\
 exp\_veldisp\_err & km/s & real & stat.error;phys.veloc.dispersion & Velocity dispersion error (exp SFH) \\
 exp\_tau & Myr & real & time.age & Age (exp SFH) \\
 exp\_tau\_err & Myr & real & stat.error;time.age & Age error (exp SFH) \\
 exp\_met & - & real & phys.abund.Z & Metallicity (exp SFH) \\
 exp\_met\_err & - & real & stat.error;phys.abund.Z & Metallicity error (exp SFH) \\
 exp\_chi2 & - & real & stat.fit.chi2 & Goodness of fit (exp SFH) \\
 ssp\_radvel & km/s & real & spect.dopplerVeloc.opt & Radial velocity (SSP) \\
 ssp\_radvel\_err & km/s & real & stat.error;spect.dopplerVeloc.opt & Radial velocity error (SSP) \\
 ssp\_veldisp & km/s & real & phys.veloc.dispersion & Velocity dispersion (SSP) \\
 ssp\_veldisp\_err & km/s & real & stat.error;phys.veloc.dispersion & Velocity dispersion error (SSP) \\
 ssp\_age & Myr & real & time.age & Age (SSP) \\
 ssp\_age\_err & Myr & real & stat.error;time.age & Age error (SSP) \\
 ssp\_met & - & real & phys.abund.Z & Metallicity (SSP) \\
 ssp\_met\_err & - & real & stat.error;phys.abund.Z & Metallicity error (SSP) \\
 ssp\_chi2 & - & real & stat.fit.chi2 & Goodness of fit (SSP) \\
 zy\_offset & mag & real & phot.mag;arith.diff & Offset applied to UKIDSS magnitudes to correct for mismatch with SDSS ones \\
 spectrum\_snr & - & real & stat.snr & Signal-to-noise ratio of SDSS spectrum at 5500A (restframe) in the 20A box \\
\hline
\end{tabular}
\end{scriptsize}
\end{table*}

\begin{table*}
\caption{Fiber magnitudes table ({\tt rcsed\_fibermags}) columns metadata and descriptions.
\label{tab_metadata_fibermags}}
\begin{tabular}{lcccp{7cm}}
Column & Units & Datatype & UCD & Description \\
\hline
 objid & - & bigint & meta.id;meta.main & SDSS ObjID (unique identifier) \\
 corrfibmag\_fuv & mag & real & phot.mag;em.UV.FUV & Galactic extinction corrected 3" aperture magnitude in GALEX $FUV$ filter \\
 corrfibmag\_nuv & mag & real & phot.mag;em.UV.NUV & Same as above for GALEX $NUV$ filter \\
 corrfibmag\_u & mag & real & phot.mag;em.opt.U & Galactic extinction corrected fiber (3" aperture) magnitude in SDSS $u$ filter \\
 corrfibmag\_g & mag & real & phot.mag;em.opt.B & Same as above for SDSS $g$ filter \\
 corrfibmag\_r & mag & real & phot.mag;em.opt.R & Same as above for SDSS $r$ filter \\
 corrfibmag\_i & mag & real & phot.mag;em.opt.I & Same as above for SDSS $i$ filter \\
 corrfibmag\_z & mag & real & phot.mag;em.opt.I & Same as above for SDSS $z$ filter \\
 corrfibmag\_y & mag & real & phot.mag;em.IR.J & Galactic extinction corrected 3" aperture magnitude in UKIDSS $Y$ filter \\
 corrfibmag\_j & mag & real & phot.mag;em.IR.J & Same as above for UKIDSS $J$ filter \\
 corrfibmag\_h & mag & real & phot.mag;em.IR.H & Same as above for UKIDSS $H$ filter \\
 corrfibmag\_k & mag & real & phot.mag;em.IR.K & Same as above for UKIDSS $K$ filter \\
 corrfibmag\_fuv\_err & mag & real & stat.error;phot.mag;em.UV.FUV & Uncertainty of corrfibmag\_fuv column \\
 corrfibmag\_nuv\_err & mag & real & stat.error;phot.mag;em.UV.NUV & Uncertainty of corrfibmag\_nuv column \\
 corrfibmag\_u\_err & mag & real & stat.error;phot.mag;em.opt.U & Uncertainty of corrfibmag\_u \\
 corrfibmag\_g\_err & mag & real & stat.error;phot.mag;em.opt.B & Uncertainty of corrfibmag\_g \\
 corrfibmag\_r\_err & mag & real & stat.error;phot.mag;em.opt.R & Uncertainty of corrfibmag\_r \\
 corrfibmag\_i\_err & mag & real & stat.error;phot.mag;em.opt.I & Uncertainty of corrfibmag\_i \\
 corrfibmag\_z\_err & mag & real & stat.error;phot.mag;em.opt.I & Uncertainty of corrfibmag\_z \\
 corrfibmag\_y\_err & mag & real & stat.error;phot.mag;em.IR.J & Uncertainty of corrfibmag\_y \\
 corrfibmag\_j\_err & mag & real & stat.error;phot.mag;em.IR.J & Uncertainty of corrfibmag\_j \\
 corrfibmag\_h\_err & mag & real & stat.error;phot.mag;em.IR.H & Uncertainty of corrfibmag\_h \\
 corrfibmag\_k\_err & mag & real & stat.error;phot.mag;em.IR.K & Uncertainty of corrfibmag\_k \\
 kcorrfib\_fuv & mag & real & arith.factor;em.UV.FUV & K-correction for 3" aperture GALEX $FUV$ magnitude \\
 kcorrfib\_nuv & mag & real & arith.factor;em.UV.NUV & Same as above for GALEX $NUV$ magnitude \\
 kcorrfib\_u & mag & real & arith.factor;em.opt.U & K-correction for fiber (3" aperture) SDSS $u$ magnitude \\
 kcorrfib\_g & mag & real & arith.factor;em.opt.B & Same as above for SDSS $g$ magnitude \\
 kcorrfib\_r & mag & real & arith.factor;em.opt.R & Same as above for SDSS $r$ magnitude \\
 kcorrfib\_i & mag & real & arith.factor;em.opt.I & Same as above for SDSS $i$ magnitude \\
 kcorrfib\_z & mag & real & arith.factor;em.opt.I & Same as above for SDSS $z$ magnitude \\
 kcorrfib\_y & mag & real & arith.factor;em.IR.J & K-correction for 3" aperture UKIDSS $Y$ magnitude \\
 kcorrfib\_j & mag & real & arith.factor;em.IR.J & Same as above for UKIDSS $J$ magnitude \\
 kcorrfib\_h & mag & real & arith.factor;em.IR.H & Same as above for UKIDSS $H$ magnitude \\
 kcorrfib\_k & mag & real & arith.factor;em.IR.K & Same as above for UKIDSS $K$ magnitude \\
\hline
\end{tabular}
\end{table*}

\begin{table*}
\caption{Gas phase metallicity table ({\tt rcsed\_gasmet}) columns metadata and descriptions.
\label{tab_metadata_gasmet}}
\begin{tabular}{lcccp{8cm}}
Column & Units & Datatype & UCD & Description \\
\hline
 id &  & bigint & meta.id;meta.main & Primary key \\
 objid &  & bigint &  &  SDSS ObjID \\
 mjd & d & integer & time.epoch & MJD of observation \\
 plate &  & smallint & meta.id & SDSS plate ID \\
 fiberid &  & smallint & meta.id & SDSS fiber ID \\
 e\_bv & mag & real & phot.color.excess & Intrinsic E(B-V) \\
 gas\_oh\_d16 &  & real & phys.abund.Z & Oxygen abundance of ionized gas (12 + log O/H) calculated using Dopita+16 calibration from Gaussian fit to emission lines\\
 gas\_oh\_d16\_err &  & real & phys.abund.Z & Error of oxygen abundance of ionized gas (12 + log O/H) calculated using Dopita+16 calibration from Gaussian fit to emission lines\\
 gas\_oh\_izi &  & real & phys.abund.Z & Oxygen abundance of ionized gas (12 + log O/H) calculated using IZI calibration from Gaussian fit to emission lines\\
 gas\_oh\_izi\_errlo &  & real & stat.error;phys.abund.Z & Lower error of oxygen abundance of ionized gas (12 + log O/H) calculated using IZI calibration from Gaussian fit to emission lines \\
 gas\_oh\_izi\_errhi &  & real & stat.error;phys.abund.Z & Upper error of oxygen abundance of ionized gas (12 + log O/H) calculated using IZI calibration from Gaussian fit to emission lines \\
 q\_izi &  & real & phys.ionizParam.rad & Ionization parameter calculated using IZI calibration from Gaussian fit to emission lines \\
 q\_izi\_errlo &  & real & stat.error;phys.ionizParam.rad & Lower error of ionization parameter calculated using IZI calibration from Gaussian fit to emission lines \\
 q\_izi\_errhi &  & real & stat.error;phys.ionizParam.rad & Upper error of ionization parameter calculated using IZI calibration from Gaussian fit to emission lines \\
\hline
\end{tabular}
\end{table*}

\begin{table*}
\caption{Gaussian fit to emission lines table ({\tt rcsed\_lines\_gauss}) columns metadata and descriptions.
\label{tab_metadata_gauss}}
\begin{tabular}{lllp{2.5cm}p{7cm}}
Column & Units & Datatype & UCD & Description \\
\hline
 id &  & bigint & meta.id; meta.main & Primary key \\
 objid &  & bigint & meta.id & SDSS ObjID \\
 mjd & d & integer & time.epoch & MJD of observation \\
 plate &  & smallint & meta.id & SDSS plate ID \\
 fiberid &  & smallint & meta.id & SDSS fiber ID \\
 forbid\_v & km/s & real & phys.veloc & Velocity measured simultaneously in all forbidden lines \\
 forbid\_v\_err & km/s & real & stat.error; phys.veloc & Uncertainty in the velocity measured simultaneously in all forbidden lines \\
 forbid\_sig & km/s & real & phys.veloc.dispersion & Velocity dispersion measured simultaneously in all forbidden lines \\
 forbid\_sig\_err & km/s & real & stat.error; phys.veloc.dispersion & Uncertainty in the velocity dispersion measured simultaneously in all forbidden lines \\
 allowed\_v & km/s & real & phys.veloc & Velocity measured simultaneously in all allowed lines \\
 allowed\_v\_err & km/s & real & stat.error; phys.veloc & Uncertainty in the velocity measured simultaneously in all allowed lines \\
 allowed\_sig & km/s & real & phys.veloc.dispersion & Velocity dispersion measured simultaneously in all allowed lines \\
 allowed\_sig\_err & km/s & real & stat.error; phys.veloc.dispersion & Uncertainty in the velocity dispersion measured simultaneously in all allowed lines \\
 chi2 &  & real & stat.fit.chi2 & Reduced goodness of fit \\
 f3727\_oii\_flx & 10$^{-17}$ erg/s/cm$^{2}$ & real & phot.flux; spect.line & Flux from Gaussian fit to continuum subtracted data of [O{\sc ii}] (3727~\AA) line \\
 f3727\_oii\_flx\_err & 10$^{-17}$ erg/s/cm$^{2}$ & real & stat.error; phot.flux; spect.line & Uncertainty in the flux from Gaussian fit to continuum subtracted data of [O{\sc ii}] (3727~\AA) line \\
 f3727\_oii\_cnt & 10$^{-17}$ erg/s/cm$^{2}$/\AA & real & phot.flux.density; spect.continuum & Continuum level at [O{\sc ii}] (3727~\AA) line center \\
 f3727\_oii\_cnt\_err & 10$^{-17}$ erg/s/cm$^{2}$/\AA & real & stat.error; phot.flux.density; spect.continuum & Uncertainty in the continuum level at [O{\sc ii}] (3727~\AA) line center \\
 f3727\_oii\_ew & \AA & real & spect.line.eqWidth & Equivalent width from Gaussian fit to continuum subtracted data of [O{\sc ii}] (3727~\AA) line \\
 f3727\_oii\_ew\_err & \AA & real & stat.error; spect.line.eqWidth & Uncertainty in the equivalent width from Gaussian fit to continuum subtracted data of [O{\sc ii}] (3727~\AA) line \\
 $\dots$& $\dots$ & $\dots$& $\dots$ & $\dots$\\
\hline
\end{tabular}
\end{table*}

\begin{table*}
\caption{Non-parametric fit to emission lines table ({\tt rcsed\_lines\_nonpar}) columns metadata and descriptions.
\label{tab_metadata_nonpar}}
\begin{tabular}{lllp{2.5cm}p{7cm}}
Column & Units & Datatype & UCD & Description \\
\hline
 id &  & bigint & meta.id; meta.main & Primary key \\
 objid &  & bigint & meta.id & SDSS ObjID \\
 mjd & d & integer & time.epoch & MJD of observation \\
 plate &  & smallint & meta.id & SDSS plate ID \\
 fiberid &  & smallint & meta.id & SDSS fiber ID \\
 forbid\_v & km/s & real & phys.veloc & Velocity measured simultaneously in all forbidden lines \\
 forbid\_sig & km/s & real & phys.veloc.dispersion & Velocity dispersion measured simultaneously in all forbidden lines \\
 allowed\_v & km/s & real & phys.veloc & Velocity measured simultaneously in all allowed lines \\
 allowed\_sig & km/s & real & phys.veloc.dispersion & Velocity dispersion measured simultaneously in all allowed lines \\
 chi2 &  & real & stat.fit.chi2 & Reduced goodness of fit \\
 f3727\_oii\_flx & 10$^{-17}$ erg/s/cm$^{2}$ & real & phot.flux; spect.line & Flux from non-parametric fit to continuum subtracted data of [O{\sc ii}] (3727~\AA) line \\
 f3727\_oii\_flx\_err & 10$^{-17}$ erg/s/cm$^{2}$ & real & stat.error; phot.flux; spect.line & Uncertainty in the flux from non-parametric fit to continuum subtracted data of [O{\sc ii}] (3727~\AA) line \\
 f3727\_oii\_cnt & 10$^{-17}$ erg/s/cm$^{2}$/\AA & real & phot.flux.density; spect.continuum & Continuum level at [O{\sc ii}] (3727~\AA) line center \\
 f3727\_oii\_cnt\_err & 10$^{-17}$ erg/s/cm$^{2}$/\AA & real & stat.error; phot.flux.density; spect.continuum & Uncertainty in the continuum level at [O{\sc ii}] (3727~\AA) line center \\
 f3727\_oii\_ew & \AA & real & spect.line.eqWidth & Equivalent width from non-parametric fit to continuum subtracted data of [O{\sc ii}] (3727~\AA) line \\
 f3727\_oii\_ew\_err & \AA & real & stat.error; spect.line.eqWidth & Uncertainty in the equivalent width from non-parametric fit to continuum subtracted data of [O{\sc ii}] (3727~\AA) line \\
 $\dots$& $\dots$ & $\dots$& $\dots$ & $\dots$\\
\hline
\end{tabular}
\end{table*}

\end{document}